\documentclass[%
 reprint,
superscriptaddress,
 amsmath,amssymb,
 aps,
]{revtex4-2}
\usepackage{graphicx}
\usepackage{dcolumn}
\usepackage{bm}
\usepackage{subcaption}
\usepackage{float}
\usepackage{xcolor}
\usepackage[hidelinks]{hyperref}

\begin{document}

\preprint{APS/123-QED}

\title{On the fluid-structure interaction of a flexible cantilever cylinder at low Reynolds numbers}

\author{Shayan Heydari}
\email[Corresponding author. Email address: ]{sheydari@mail.ubc.ca}
\affiliation{Department of Mechanical Engineering, The University of British Columbia, Vancouver, BC Canada V6T 1Z4}

\author{Neelesh A. Patankar}
\affiliation{Department of Mechanical Engineering, Northwestern University, 2145 Sheridan Road, Evanston, Illinois 60208, USA}

\author{Mitra J. Z. Hartmann}
\affiliation{Department of Mechanical Engineering, Northwestern University, 2145 Sheridan Road, Evanston, Illinois 60208, USA}

\author{Rajeev K. Jaiman}
\email[Email address: ]{rjaiman@mech.ubc.ca}
\affiliation{Department of Mechanical Engineering, The University of British Columbia, Vancouver, BC Canada V6T 1Z4}

\date{\today}

\begin{abstract}
We present a numerical study to investigate the fluid-structure interaction of a flexible circular cantilever cylinder in a uniform cross-flow. We employ a fully-coupled fluid-structure solver based on the three-dimensional Navier-Stokes equations and the Euler-Bernoulli beam theory. We examine the dynamics of the cylinder for a wide range of reduced velocities ($U^*$), mass ratios ($m^*$), and Reynolds numbers ($Re$). Of particular interest is to explore the possibility of flow-induced vibrations in a slender cantilever cylinder of aspect ratio $AR=100$ at laminar subcritical $Re$ regime (i.e., no periodic vortex shedding). We assess the extent to which such a flexible cylindrical beam can sustain flow-induced vibrations and characterize the contribution of the beam's flexibility to the stability of the wake at low $Re$. We show that when certain conditions are satisfied, the flexible cantilever cylinder undergoes sustained large-amplitude vibrations. The frequency of the oscillations is found to match the frequency of the periodic fluid forces for a particular range of system parameters. In this range, the frequency of the transverse vibrations is shown to match the first-mode natural frequency of the cylinder, indicating the existence of the lock-in phenomenon. The range of the lock-in regime is shown to have a strong dependence on $Re$ and $m^*$. We discover that unlike the steady wake behind a stationary rigid cylinder, the wake of a low mass ratio flexible cantilever cylinder could lose its stability in the lock-in regime at Reynolds numbers as low as $Re=22$. A combined VIV-galloping type instability is shown to be the possible cause of the wake instability at this $Re$ regime. These findings attempt to generalize our understanding of the flow-induced vibrations in flexible cantilever structures and can have a profound impact on the development of novel flow-measurement sensors.
\end{abstract}

\maketitle
\section{\label{sec:introduction}Introduction\protect}
Flow-induced vibrations (FIVs) have significant consequences and are essential to predict in numerous fields, such as marine/offshore, civil, biomedical, and aerospace engineering. Considerable research has been done in recent decades to characterize the underlying mechanism and explore the practical aspects of flow-induced vibrations in a wide range of domains, including vibration control~\cite{Law2017,Chizfahm2021,PhysRevLett.100.204501}, energy harvesting~\cite{PhysRevApplied.3.014009,Chizfahm2018,Hobbs2012,Lee2019,Gurugubelli2015}, and sensing~\cite{Gul2018, Scharff2019}. 
%
The phenomenon of flow-induced vibration in bluff bodies has received special attention in the literature due to complex vortex dynamics and nonlinear physics involved in the interaction of a bluff body and fluid flow. In this regard, the flow-induced vibration of an elastically-mounted rigid cylinder has served as a prototypical model for both experimental and numerical studies~\cite{blevins}. It has been shown that asymmetric vortex shedding from the wake of an elastically-mounted rigid cylinder exerts unsteady transverse loads that could lead to sustained large-amplitude vibrations called vortex-induced vibrations (VIVs)~\cite{Khalak1999}. VIVs are characterized by a frequency match between the frequency of the periodic vortex shedding and the vibration frequency of the cylinder~\cite{Khalak1999,Sarpkaya1979}. When the natural frequency of the cylinder is close to the vortex shedding frequency, the VIV phenomenon results in a complex evolution of the shedding frequency, which deviates from the Strouhal frequency of its stationary counterpart. In this frequency regime, the vortex formation locks on to the natural frequency of the structure, which in turn creates a strong coupling between the cylinder and fluid flow~\cite{Khalak1999}. 
%
Several studies have shown that the peak vibration amplitude of an elastically-mounted rigid cylinder, with only one degree-of-freedom in the transverse direction, is approximately $O(D)$~\cite{Williamson2004}, where $D$ denotes the cylinder diameter. The magnitude of the peak vibration amplitude is known to be a function of fluid and structural parameters, such as Reynolds number and mass-damping ratio~\cite{Bearman2011}, and has been shown to have a slightly higher value for a two-degree-of-freedom cylinder~\cite{jauvtis_williamson_2004}. Comprehensive reviews regarding the VIV of elastically-mounted rigid cylinders could be found in Refs.~\cite{Williamson2004,govardhan_williamson_2006,Williamson2008,Sarpkaya2004,Bearman2011}.

More recently, studies have focused on the dynamic response of flexible slender structures at high Reynolds numbers~\cite{Willden,Trim,Vandiver_2009} to give new physical insight into the phenomenon of vortex-induced vibrations. 
Due to the complex interaction of nonlinear wake dynamics with numerous flexible modes, the VIV modeling and prediction poses serious challenges for long flexible structures.
For example, studies on thin risers have found that ocean currents excite several vibration modes and frequencies along the span of a riser during VIVs~\cite{Bourguet2011,Joshi2017, Franzini2018}. In a short-term perspective, VIV effects can lead to drag amplification and large dynamic bending stresses. These large-amplitude vibrations lead to fatigue failure in slender structures and marine risers in the long term if not controlled properly~\cite{BAARHOLM2006109}. An Experimental study on the VIV of a flexible cantilever cylinder in the laminar flow regime has found some distinct differences between the dynamics of a flexible cantilever cylinder and a flexible riser in the VIV regime. A flexible pinned-pinned beam, such as a marine riser, has been shown to vibrate at monotonically increasing frequencies with each eigenmode gradually growing in modal weight as the reduced velocity is increased~\cite{Chaplin2005}. However, for a flexible cantilever cylinder, although higher modes are observed at higher reduced velocities, the cylinder has been shown to oscillate with only one vibration eigenmode during VIVs~\cite{Shang2014}. In line with the works done in the field of vortex-induced vibrations, in our current work, we examine the dynamic response of a flexible cantilever cylinder at low Reynolds numbers to give new insight into the topic of flow-induced vibrations in flexible slender structures.

Our interest in studying the dynamics of a flexible cantilever cylinder at low Reynolds numbers originates from the intriguing problem of sensing through whiskers in some mammals, such as rats and seals. Experimental studies on the mechanical response of isolated rat vibrissae (whiskers) to low-speed airflow have revealed that air currents of magnitude $0.5$ to $5.6$ m/s, typically found in natural environments, generate significant vibrissal motion that carries information about the direction and magnitude of the airflow ~\cite{Yu2019,Yu2016}. More interestingly, behavioral experiments have shown that rats use the information from their whiskers to localize airflow sources~\cite{Yue1600716}. To characterize the geometry of a rat's tapered conical whisker, two parameters, namely arc length $S$ and base diameter $D_\mathrm{b}$, are defined. Typically, in a rat's mystacial pad, the ratio of a whisker's arc length $S$ to base diameter $D_\mathrm{b}$ is between $100<S/D_\mathrm{b}<400$ and the wind speeds translate into Reynolds numbers $<50$, based on $D_\mathrm{b}$ of the whisker~\cite{Yu2019}.

In addition to rats, harbor seals, for example, have been shown to use their highly sensitive undulated whiskers to sense hydrodynamic information of water flows and detect fluid structures without auditory or visual cues~\cite{Dehnhardt102}. An experimental study on a model of a seal whisker has shown that a whisker in the wake of a stationary rigid cylinder undergoes large-amplitude oscillations, with the frequency of the oscillations being close to the shedding frequency of the upstream wake~\cite{Beem2015}. These oscillations are known to help seals detect upstream wakes and estimate the size and shape of the wake-generating body~\cite{Beem2015}. 

Although considerable research has been done to understand the problem of sensing through whiskers, there have been limited studies on the mechanism behind the dynamics of whiskers in fluid flow. In particular, the oscillatory response of a whisker at laminar subcritical Reynolds numbers, i.e., $Re < Re_\mathrm{cr}\approx45$~\cite{Park2016,Yao2017}, is not well understood. 
In our current work, we investigate fluid-structure interaction of a flexible cantilever cylinder, as a simplified model of a whisker, to help answer two specific questions: (i) can we observe sustained vibrations in the flexible cantilever cylinder at subcritical $Re$ with laminar wake flow, and (ii) what is the relationship between the cylinder dynamics and stability of the wake at this $Re$ regime? Understanding the underlying fluid-structure dynamics of a flexible cantilever cylinder, inspired by the dynamics of whiskers, is of vital importance for developing novel flow-measurement sensors~\cite{Gul2018} and brings us one step closer towards a complete mapping of the sensing properties of whiskers.

The key non-dimensional parameters involved in the fluid-structure interaction of the flexible cantilever cylinder are mass ratio $m^\mathrm{*}$, Reynolds number $Re$, and reduced velocity $U^\mathrm{*}$ defined as
\begin{align}
	m^\mathrm{*} = \frac{4 m}{{\pi D^{2}\rho^\mathrm{f}}}, \qquad
	Re = \frac{\rho^\mathrm{f}U_\mathrm{0}D}{\mu^\mathrm{f}}, \qquad
	U^\mathrm{*} &= \frac{U_\mathrm0}{f_\mathrm{n}D},
\end{align}
where $m$ is the mass per unit length of the cylinder, $D$ is the cylinder diameter, $\rho^\mathrm{f}$ and $\mu^\mathrm{f}$ are the density and dynamic viscosity of the fluid, respectively, $U_0$ is the magnitude of the uniform flow velocity, and $f_\mathrm{n}$ is the natural frequency of the first mode of vibration. The non-dimensional parameters studied in our current work are within $20\leq Re\leq40$, $U^*\in[2,19]$ and $1\leq m^*\leq1000$, which cover a practical range of values in the laminar subcritical $Re$ regime.

\begin{figure}[b]
\includegraphics[width=1\linewidth]{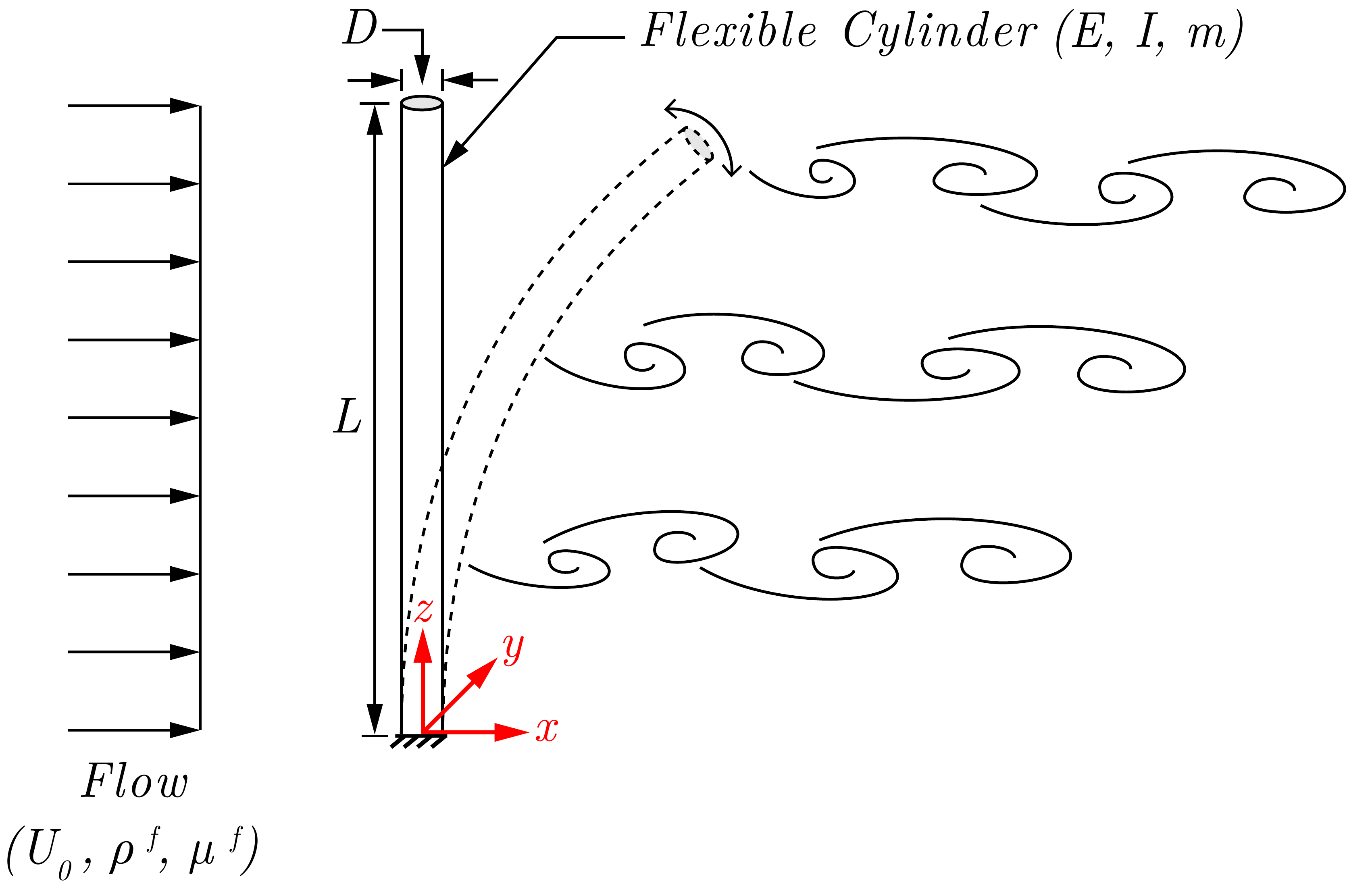}
\caption{\label{Schematic_Problem_Statement}Schematic of a flexible cantilever cylinder of diameter $D$ and length $L$ interacting with a uniform flow of velocity $U_0$.}
\end{figure}

A schematic of the flexible cantilever cylinder is given in Fig.~\ref{Schematic_Problem_Statement}. The cylinder is connected to a fixed support at $z=0$.
The Young's modulus and second moment of area of the cylinder are denoted by $E$ and $I$, respectively. As shown in Fig.~\ref{Schematic_Problem_Statement}, due to fluid forces acting on the cylinder, it initially deforms in the streamwise direction. Also, depending on the system parameters, the cylinder could exhibit an unsteady dynamic response, resulting in periodic vortex shedding patterns in the wake region. A systematic analysis of the dynamic response of the cylinder is conducted in our current work, where we use high-fidelity numerical simulations to examine the fluid-structure interaction of the cylinder for a broad range of system parameters.

The content of the paper is structured as follows. The governing equations for modeling the cylinder dynamics and the coupling strategy between the fluid and structural solvers are discussed in Section~\ref{sec:numericalMethodology}. In addition, we provide the results for the grid convergence study at the end of this section. In section~\ref{sec:results} we cover the results of our study and discuss the dynamic response characteristics of the cylinder in detail. Finally, we finish the paper with a conclusion in section~\ref{sec:conclusions}.
\section{\label{sec:numericalMethodology} Numerical Methodology}
\begin{figure*}
\includegraphics[scale=0.3]{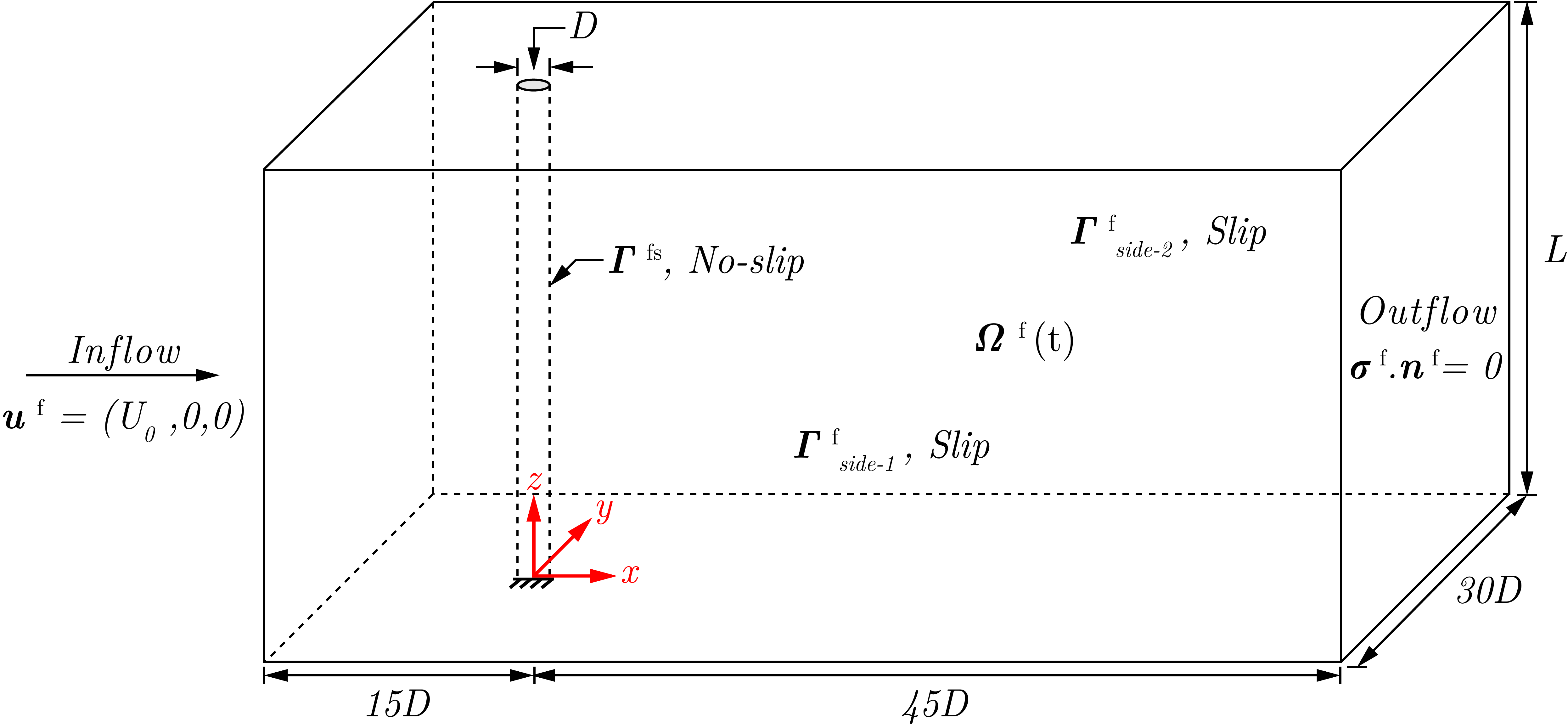}
\caption{\label{Beam_Schematic}Schematic of the computational domain with details of the boundary conditions.}
\end{figure*}

This section presents a three-dimensional numerical framework for studying the fluid-structure interaction of the flexible cantilever cylinder with an incompressible viscous flow. To model the coupled dynamics of the cylinder, we employ a three-dimensional computational domain as shown in Fig.~\ref{Beam_Schematic}. The cylinder is placed at an offset distance of $15D$ and $45D$ from the inflow and outflow surfaces, respectively. Fixed structural support is imposed at one end of the cylinder ($z=0$), and the no-slip boundary condition is applied at the fluid-structure interface $\Gamma^\mathrm{fs}$. The size of the computational domain is $60D\times30D\times L$ where a uniform velocity of $\boldsymbol{u}^\mathrm{f} = (U_0,0,0)$ with a magnitude of $U_0$ in the x-direction is given at the inflow surface. The slip boundary condition is applied to the side surfaces $\boldsymbol\Gamma^{\mathrm{f}}_{side-1}$ and $\boldsymbol\Gamma^{\mathrm{f}}_{side-2}$, and the traction-free boundary condition, given by $ \boldsymbol{\sigma}^{\mathrm{f}}.\boldsymbol{n}^\mathrm{f} = 0$, is specified at the outflow surface.

In the following, we discuss the governing equations for modeling the dynamics of the flexible cantilever cylinder and present the strategy implemented to couple the fluid and structural solvers.
\subsection{Governing equations}
We consider the three-dimensional incompressible Navier-Stokes equations coupled with the Euler-Bernoulli beam theory to examine the coupled dynamics of the flexible cantilever cylinder. We formulate the governing equation for the Euler-Bernoulli beam in a Lagrangian reference frame and take a body-fitted moving boundary approach based on the arbitrary Lagrangian-Eulerian (ALE) description~\cite{Hughes1981} to formulate the unsteady Navier-Stokes equations for the viscous incompressible fluid. The body-fitted treatment of the fluid-structure interface through the ALE description of the flow field provides accurate modeling of the boundary layer over the deformable surface of the structure.

\subsubsection{Navier–Stokes equations for a moving-boundary problem}
The unsteady Navier-Stokes equations for a viscous incompressible fluid flow in an arbitrary Lagrangian-Eulerian reference frame on the fluid domain $\Omega^\mathrm{f}(t)$ are
\begin{align} \label{DNS}
	\rho^\mathrm{f}\frac{\partial \boldsymbol{u}^\mathrm{f}}{\partial t}\bigg|_{\hat{x}^\mathrm{f}} + \rho^\mathrm{f}(\boldsymbol{u}^\mathrm{f} - {{\boldsymbol{u}^\mathrm{m}}})\cdot\nabla\boldsymbol{u}^\mathrm{f} &= \nabla\cdot \boldsymbol{\sigma}^\mathrm{f} + \boldsymbol{b}^\mathrm{f}\ \ \ \mathrm{on\ \ \Omega^\mathrm{f}(t)},\\
	\nabla\cdot\boldsymbol{u}^\mathrm{f} &= 0\ \ \ \mathrm{on\ \ \Omega^\mathrm{f}(t)},
\end{align}
where $\boldsymbol{u}^\mathrm{f} = \boldsymbol{u}^\mathrm{f}(\boldsymbol{x}^\mathrm{f},t)$ and $\boldsymbol{u}^\mathrm{m}=\boldsymbol{u}^\mathrm{m}(\boldsymbol{x}^\mathrm{f},t)$ denote the fluid and mesh velocities defined for each spatial point $\boldsymbol{x}^\mathrm{f} \in \Omega^\mathrm{f}(t)$ respectively,  $\boldsymbol{b}^\mathrm{f}$ is the body force applied to the fluid and $\boldsymbol{\sigma}^\mathrm{f}$ is the Cauchy stress tensor for a Newtonian fluid, given as
\begin{align}
	\boldsymbol{\sigma}^\mathrm{f} = -p\boldsymbol{I} + \mu^\mathrm{f}( \nabla\boldsymbol{u}^\mathrm{f} + (\nabla\boldsymbol{u}^\mathrm{f})^T),
\end{align}
where $p$ denotes the fluid pressure, and $\mu^\mathrm{f}$ is the dynamic viscosity of the fluid. The first term in Eq.~(\ref{DNS}) represents the partial derivative of $\boldsymbol{u}^\mathrm{f}$ with respect to time while the ALE referential coordinate $\hat{x}^\mathrm{f}$ is kept fixed.

The fluid forcing acting on the beam's surface is calculated by integrating the surface traction at the first layer of the elements located on the fluid-structure interface. The instantaneous coefficients of lift and drag forces are quantified as
\begin{align}
	C_\mathrm{L} = \frac{1}{\frac{1}{2}\rho^\mathrm{f}U_\mathrm{0}^\mathrm{2}DL}\int_{\Gamma^\mathrm{fs}} (\boldsymbol{\sigma}^\mathrm{f}\cdot \boldsymbol{n})\cdot \boldsymbol{n}_\mathrm{y} \mathrm{d\Gamma}, \\
C_\mathrm{D} = \frac{1}{\frac{1}{2}\rho^\mathrm{f}U_\mathrm{0}^\mathrm{2}DL}\int_{\Gamma^\mathrm{fs}} (\boldsymbol{\sigma}^\mathrm{f}\cdot \boldsymbol{n})\cdot \boldsymbol{n}_\mathrm{x} \mathrm{d\Gamma},
\end{align}
where $\boldsymbol{n}_\mathrm{x}$ and $\boldsymbol{n}_\mathrm{y}$ are the Cartesian components of the unit outward normal vector $\boldsymbol{n}$. 
In the next section, we present the equation for modeling the structural dynamics of the flexible cantilever cylinder using the Euler-Bernoulli beam theory.
\subsubsection{\label{subsec:Euler-Bernoulli beam eqn}Euler-Bernoulli beam theory for a flexible structure}
We consider the flexible cantilever cylinder as a slender structure with relatively small lateral motions. Therefore, the Euler Bernoulli beam theory can be applied to model its dynamic response. Let $\Omega^\mathrm{s}$ be the structural domain consisting of structure coordinates $\boldsymbol{x}^\mathrm{s} = (x,y,z)$.
We solve the transverse displacements $\boldsymbol{w}^\mathrm{s}(z,t)$ using the Euler-Bernoulli beam equation excited by the distributed unsteady fluid force per unit length $\boldsymbol{f}^\mathrm{s}$. The motion of the flexible cantilever cylinder is governed by the fluid forces and involves integrating pressure and shear stress effects on the cylinder surface. Neglecting the damping and shear effects, we take the equation of motion for the flexible cantilever cylinder as:
\begin{align}\label{eq:beam_eqn}
	{m}\frac{\partial^2 \boldsymbol{w}^\mathrm{s}(z,t)}{\partial t^2} + EI\frac{\partial^4 \boldsymbol{w}^\mathrm{s}(z,t)}{\partial z^4}  = \boldsymbol{f}^\mathrm{s}(z,t),
\end{align}  
where $m=\rho^\mathrm{s} A$ is the mass per unit length of the cylinder, with $A$ being the cross-sectional area of the cylinder. Under the cantilever (clamped-free) configuration, the boundary conditions at the clamped end of the cylinder are given as:
\begin{align}
	\boldsymbol{w}^\mathrm{s}(z,t)|_{z=0} &= 0,\qquad
	\frac{\partial\boldsymbol{w}^\mathrm{s}(z,t)}{\partial z}\bigg|_{z=0} = 0.
\end{align}
To solve Eq. (\ref{eq:beam_eqn}), we consider a mode superposition approach for the dynamic response of the cylinder. The $\mathrm{n}^\mathrm{th}$ mode natural frequency of the flexible cantilever cylinder is given by 
\begin{align} \label{eq:natural_frequency}
	f_\mathrm{n} = \frac{{\lambda_\mathrm{n}}^2}{2\pi L^2}\sqrt{ \frac{EI}{m+m_\mathrm{a}} },
\end{align}
where $\mathrm{n}$ is the mode number, $m_\mathrm{a}$ is the added mass of the fluid per unit length defined as $m_\mathrm{a} = \pi D^{2}\rho^\mathrm{f}/4$ and $\lambda_\mathrm{n}$ is the dimensionless frequency parameter for the $\mathrm{n}^\mathrm{th}$ mode of vibration. The $\lambda_\mathrm{n}$ values are given in Table~\ref{tab:modalParameters}. The modal parameters are based on the values reported in Ref.~\cite{Blevins2016} for flexible cantilever beams of constant cross section.

\begin{table}[b]
\caption{\label{tab:modalParameters} Modal parameters for flexible cantilever beams of constant cross section~\cite{Blevins2016}.}
\begin{ruledtabular}
\begin{tabular}{ccc}
Mode number, $\mathrm{n}$&$\lambda_\mathrm{n}$&
\multicolumn{1}{c}{\textrm{$\sigma_\mathrm{n}$}}\\
\hline
1&1.87510407&0.734095514\\
2&4.69409113&1.018467319\\
3&7.85475744&0.999224497\\
4&10.99554073&1.000033553\\
5&14.13716839&$\approx$1\\
$\geq6$&(2$\mathrm{n}$-1)$\pi$/2&$\approx$1\\
\end{tabular}
\end{ruledtabular}
\end{table}

The cylinder motion is solved using simple linear vibration analysis. The displacements from the mean position of the cylinder are assumed to be small and characterized based on the normal modes of the vibration found using an eigenvalue analysis. The mode shapes of the cantilever cylinder are taken as the sums of sine, cosine, sinh, and cosh functions of $\lambda_\mathrm{n} z/L$ written as:
\begin{eqnarray} \label{eq:modal_shape}
S^\mathrm{n}\left( z \right) &=& \cosh \left( \frac{\lambda_\mathrm{n}z} {L} \right) - \cos \left( \frac{\lambda_\mathrm{n} z} {L} \right) - \sigma_\mathrm{n} \sinh \left( \frac{\lambda_\mathrm{n} z} {L} \right) \nonumber \\
& &+ \sigma_\mathrm{n} \sin \left( \frac{\lambda_\mathrm{n} z} {L} \right),
\end{eqnarray}
where $S^\mathrm{n}$ denotes the mode shape associated with the $\mathrm{n}^\mathrm{th}$ mode of vibration and $\sigma_\mathrm{n}$ is the non-dimensional parameter dependent on the mode number (see Table~\ref{tab:modalParameters} for $\sigma_\mathrm{n}$ values).
\subsubsection{Treatment of the fluid-structure interface}
We need to satisfy the continuity of velocity and traction at the fluid-structure interface. Let $\Gamma^\mathrm{fs} = \partial \Omega^\mathrm{f}(0) \cap \partial \Omega^\mathrm{s}$ be the fluid-structure interface at $t=0$ and $\Gamma^\mathrm{fs}(t) = \boldsymbol{\varphi}^\mathrm{s}(\Gamma^\mathrm{fs},t)$  be the interface at time $t$. The required conditions to be satisfied are as follows:
\begin{align}
	\boldsymbol{u}^\mathrm{f}(\boldsymbol{\varphi}^\mathrm{s}(\boldsymbol{x}^\mathrm{s}_0,t),t) = \boldsymbol{u}^\mathrm{s}(\boldsymbol{x}^\mathrm{s}_0,t), \\
	\int_{\boldsymbol{\varphi}^\mathrm{s}(\gamma,t)} \boldsymbol{\sigma}^\mathrm{f}(\boldsymbol{x}^\mathrm{f},t)\cdot \boldsymbol{n} \mathrm{d\Gamma}(\boldsymbol{x}^\mathrm{f}) + \int_\gamma  \boldsymbol{t}^\mathrm{s} \mathrm{d}\Gamma = 0,
\end{align}
where $\boldsymbol{\varphi}^\mathrm{s}$ denotes the position vector that maps the initial position $\boldsymbol{x}^\mathrm{s}_0$ of the flexible cantilever cylinder to its position at time $t$, i.e., $\boldsymbol{\varphi}^\mathrm{s}(\boldsymbol{x}^\mathrm{s},t) = \boldsymbol{x}^\mathrm{s}_0 + \boldsymbol{w}^\mathrm{s}(\boldsymbol{x}^\mathrm{s},t)$, $\boldsymbol{t}^\mathrm{s}$ is the fluid traction vector relating to the fluid forcing as 
$\boldsymbol{f}^\mathrm{s}(z,t) = 
\int_{\Gamma^\mathrm{fs}} \boldsymbol{t}^\mathrm{s} \mathrm{d}\Gamma$, and  $\boldsymbol{u}^\mathrm{s}$ is the velocity of the structure at time $t$ given by $\boldsymbol{u}^\mathrm{s} = \partial\boldsymbol{\varphi}^\mathrm{s}/\partial t$. Here, $\boldsymbol{n}$ is the outer normal to the fluid-structure interface, $\gamma$ is any part of the interface $\Gamma^\mathrm{fs}$ in the reference configuration, $\mathrm{d\Gamma}$ is the differential surface area and $\boldsymbol{\varphi}^\mathrm{s}(\gamma,t)$ is the corresponding fluid part at time $t$. The above conditions are satisfied such that the fluid velocity is exactly equal to the velocity of the structure at the fluid-structure interface. 

To couple the fluid and structure equations, we use a nonlinear partitioned iterative approach based on the nonlinear iterative force correction (NIFC) scheme described in \cite{Jaiman2016,Jaiman2016_2}. At each time step, the fluid traction applied to the surface of the cylinder is projected onto the eigenvectors to find the values of the generalized modal forces. The projected modal forces are then used to determine the modal amplitudes and displacements for the next time step. 

To account for the changes in the cylinder geometry, we explicitly control the motion of each mesh node while satisfying the kinematic consistency of the discretized interface. The movement of the internal finite element nodes is chosen such that the mesh quality does not deteriorate as the displacements of the cylinder become large. For this purpose, we assume the fluid mesh to represent a hyperelastic solid model. In addition, a standard Lagrangian finite element technique is used to adapt the mesh to the new geometry of the domain.
\subsection{Grid convergence study}
The coupled dynamics of the flexible cantilever cylinder is examined through a numerical framework that has been verified and validated extensively for FSI problems in an earlier study~\cite{Joshi2017}. Thus, we proceed with the grid convergence study here. We discretize the computational domain into unstructured hexahedral finite element grids with a boundary layer mesh around the flexible cantilever cylinder. 

\begin{figure*}
\begin{subfigure}{0.25\textwidth}
\centering
\includegraphics[width=0.9\linewidth]{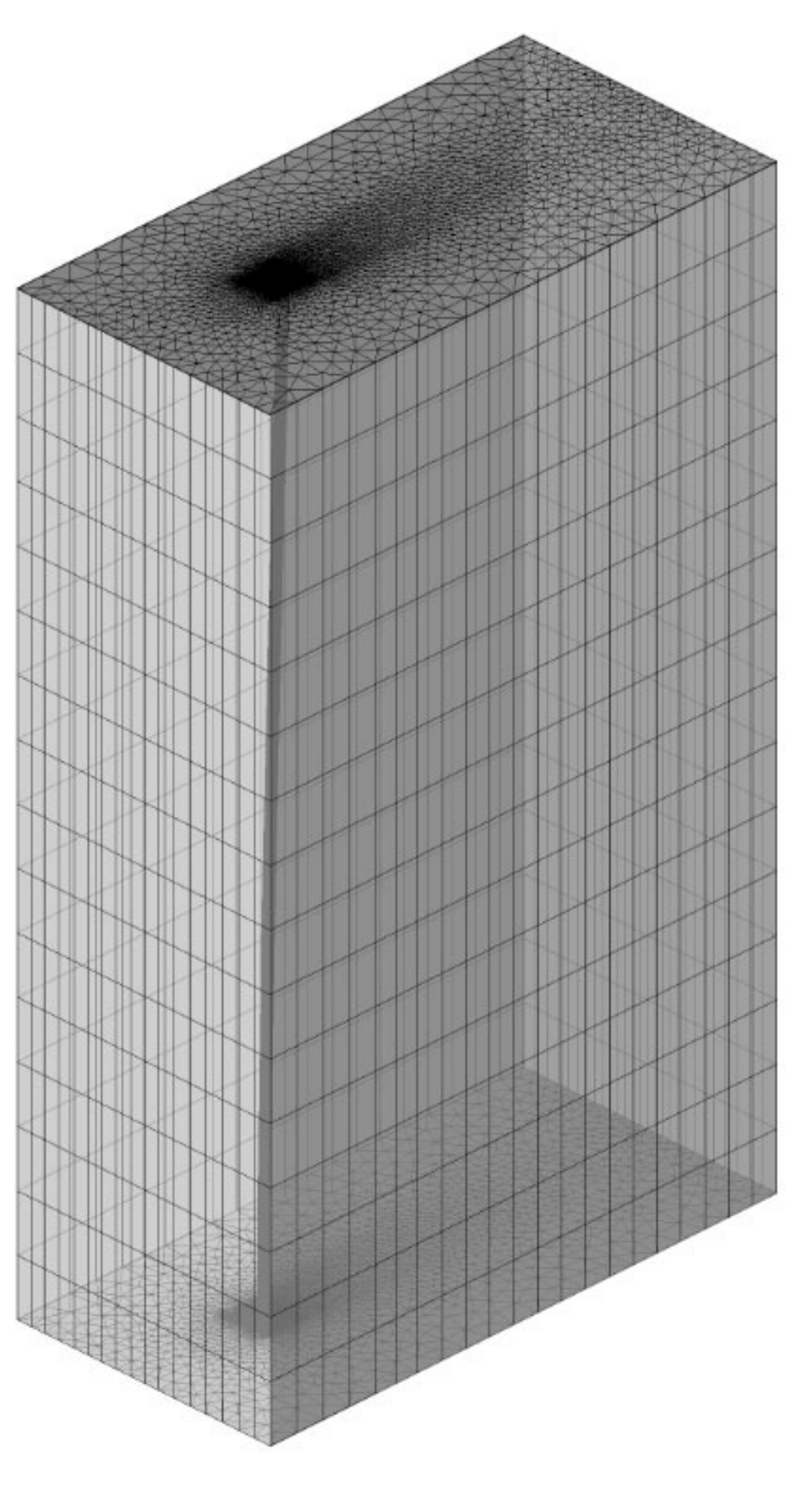}
\caption{}
\label{isoMesh}
\end{subfigure}%
\begin{subfigure}{0.75\textwidth}
\centering
\includegraphics[width=1\linewidth]{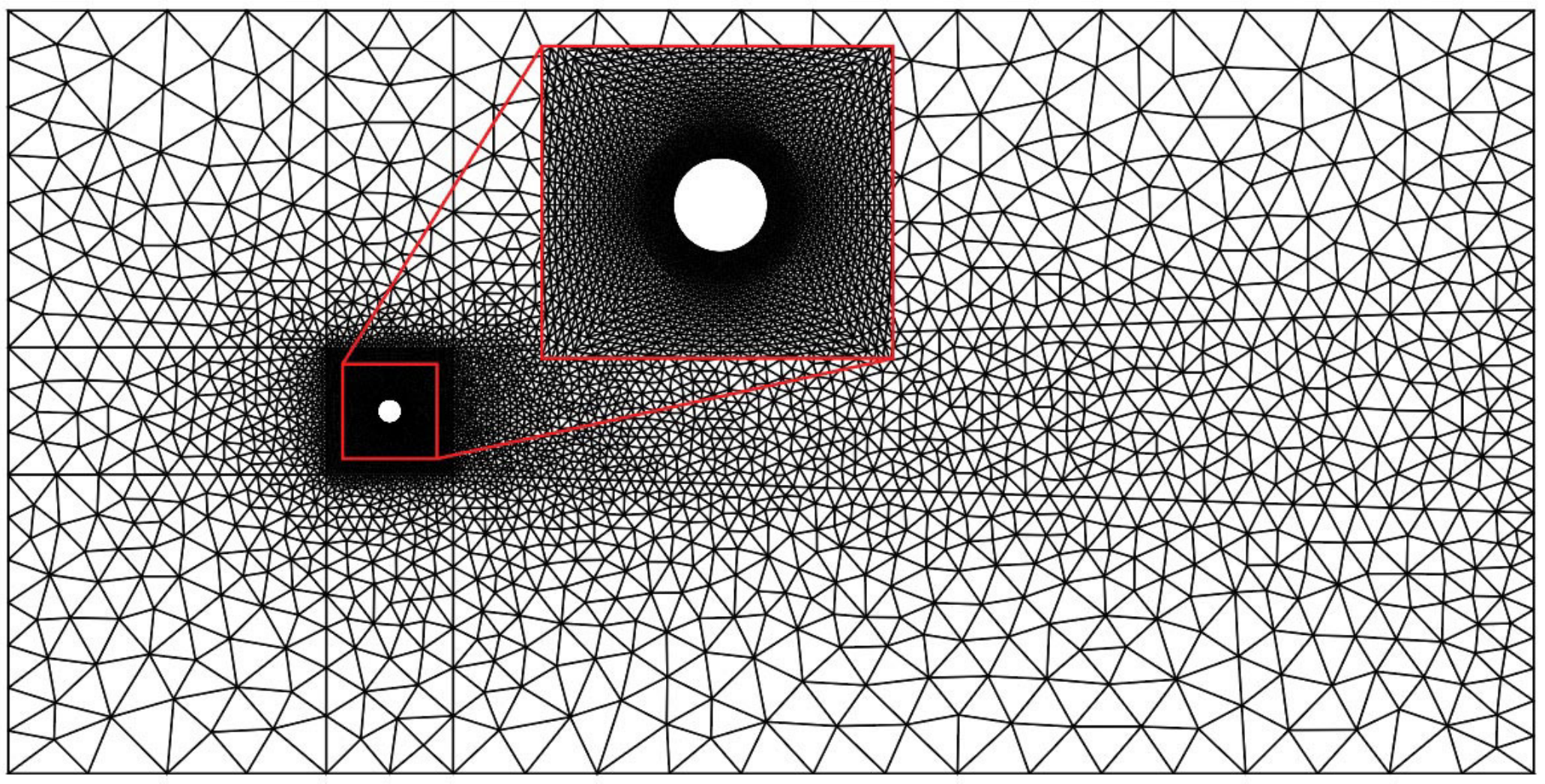}
\caption{}
\label{XYMesh}
\end{subfigure}
\caption{Computational finite element grid for the M2 mesh: (a) isometric view of the discretized computational domain; (b) representative z-plane slice of the unstructured grid with a closeup view of the boundary layer mesh.}
\label{mesh}
\end{figure*}

We start with a relatively coarse grid denoted by M1 and successively increase the number of elements by approximately a factor of 2 to achieve the M2 and M3 meshes. An isometric view of the discretized domain and a $z$-plane slice of the unstructured grid for the M2 mesh is given in Fig.~\ref{mesh}.

For the grid convergence study, we have examined the dynamics of the cylinder at $Re = 40$, $m^\mathrm{*} = 1$, and $U^\mathrm{*} = 11$. Grid convergence errors are calculated by taking the finest mesh, M3, as the reference case. The values of the frequency ratio ($f_\mathrm{y}/f_\mathrm{n}$), mean streamwise deformation ($\overline{A_\mathrm{x}}/D$), root-mean-square (rms) of the dimensionless transverse vibration amplitude ($A_\mathrm{y}^{rms}/D$), and the force coefficients ($\overline{C_\mathrm{D}}$ and $C_\mathrm{L}^{rms}$) are given in Table~\ref{tab:meshConvergence}. 

According to Table~\ref{tab:meshConvergence}, the relative errors using the M2 mesh are less than $2\%$; therefore, the M2 mesh is chosen as the suitable grid for our present study. The results mentioned in Table~\ref{tab:meshConvergence} are for a computational domain with 16 layers in the spanwise direction. After doing an independent grid convergence study on the number of spanwise layers, ranging from 8 to 64, we found that 16 layers are adequate to capture the essential three-dimensional features of the fluid-structure system.

\begin{table*}
\caption{\label{tab:meshConvergence}%
Grid convergence study results for the flexible cantilever cylinder interacting with a uniform cross-flow at $Re = 40$, $m^\mathrm{*} = 1$, and $U^\mathrm{*} = 11$.
}
\begin{ruledtabular}
\begin{tabular}{lccc}
 &
\textrm{M1}&
\textrm{M2}&
\textrm{M3}\\
\colrule
Number of nodes & 142290 & 285396 & 564672\\
Number of elements & 271458 & 547470 & 1086591\\
Time-step size $\Delta\mathrm{t}$ & 0.001 & 0.001 & 0.001\\
Frequency ratio $f_{y}/f_\mathrm{n}$ & 1.2522 & 1.2522 & 1.2522\\
Mean streamwise deformation $\overline{A_\mathrm{x}}/D$  & 2.6000 (0.26\%) & 2.5966 (0.13\%) & 2.5933\\
rms of transverse vibration amplitude $A_\mathrm{y}^{rms}/D$   & 0.3188 (1.98\%) & 0.3134 (0.25\%) & 0.3126\\
Mean drag coefficient $\overline{C_\mathrm{D}}$ & 1.6958 (0.37\%) & 1.6917 (0.13\%) & 1.6895\\
rms of lift coefficient $C_\mathrm{L}^{rms}$ & 0.0197 (14.53\%) & 0.0175 (1.74\%) & 0.0172\\
\end{tabular}
\end{ruledtabular}
\end{table*}

\section{\label{sec:results} Results and discussion}
In this section, we present our results for the fluid-structure interaction of the flexible cantilever cylinder for $20\leq Re\leq40$, $U^*\in[2,19]$ and $1\leq m^*\leq 1000$. In addition, we discuss the wake dynamics for the range of studied parameters. Finally, we relate our findings to real-world observations regarding the oscillatory motion of whiskers in laminar fluid flow.
\subsection{\label{subsec:responseCharacteristics} Response characteristics}
\begin{figure*}
\centering
\includegraphics[width=0.5\linewidth]{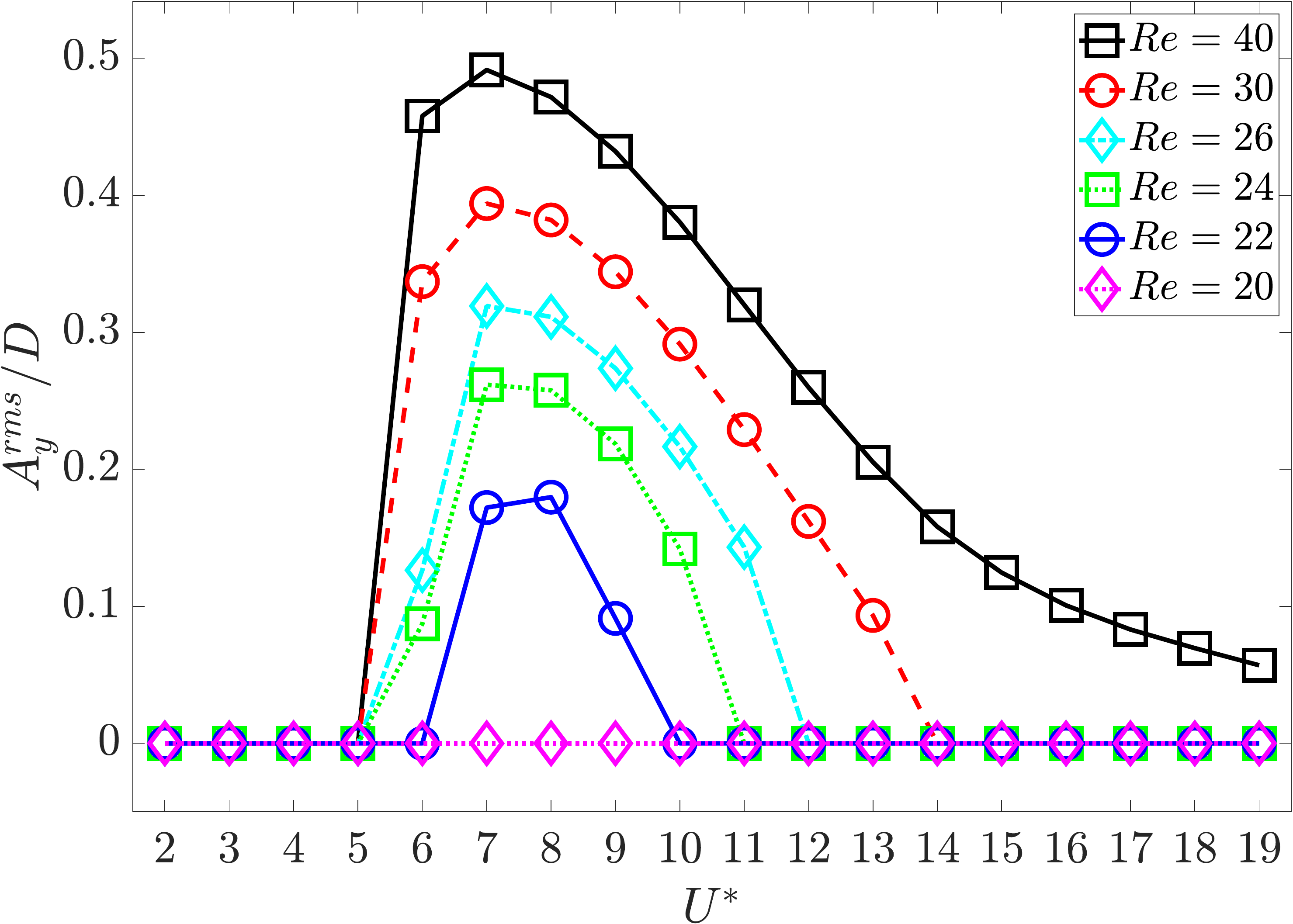}
\caption{\label{Response_Characteristics}Root-mean-square (rms) value of the dimensionless transverse vibration amplitude $A_\mathrm{y}^{rms}/D$ at $z/L=1$ as a function of $U^*$ at $m^*=1$ for $20\leq Re \leq 40$.}
\end{figure*}

We first present the response characteristics of the flexible cantilever cylinder at $m^*=1$ for $20\leq Re\leq 40$ and $U^*\in[2,19]$. The root-mean-square (rms) values of the dimensionless transverse vibration amplitude $A_\mathrm{y}^{rms}/D$ at the tip of the cylinder ($z/L=1$) is given in Fig.~\ref{Response_Characteristics}. We find that at $Re=20$, the cylinder remains in its steady deflected position, i.e., $A_\mathrm{y}^{rms} = 0$, within the range of studied $U^*$. This steady response is also observed at $Re=22$ for $U^*\leq 6$ and at $24 \leq Re\leq 40$ for $U^*\leq 5$ (see Fig.~\ref{Response_Characteristics}). However, there is a particular range of $U^*$ within which the cylinder is shown to undergo sustained vibrations for $22\leq Re\leq 40$. We observe that the peak of the transverse vibration amplitude in this range is within $U^*\in[7,8]$. As shown in Fig.~\ref{Response_Characteristics}, at $Re=22$, the peak of the $A_\mathrm{y}^{rms}/D$ is at $U^*=8$ with a magnitude of approximately $0.18$. However, at higher $Re$, the peak of the vibration amplitude shifts to $U^*=7$, where the oscillations are shown to grow in magnitude as $Re$ is increased. For instance, the peak of the $A_\mathrm{y}^{rms}/D$ has a magnitude of approximately $0.26$ at $Re=24$, whereas at $Re=40$, the maximum $A_\mathrm{y}^{rms}/D$ is approximately equal to $0.49$. We find that at $Re=40$, the cylinder experiences sustained oscillations for reduced velocities between $U^*\in[6,19]$; however, for lower $Re$, the oscillations are present for a narrower range of $U^*$. A more broadband oscillatory response with respect to $U^*$ at higher $Re$ is due to larger inertial fluid forces that overcome the viscous damping.

\begin{figure*}
\begin{subfigure}{0.5\textwidth}
\centering
\includegraphics[width=0.82\linewidth]{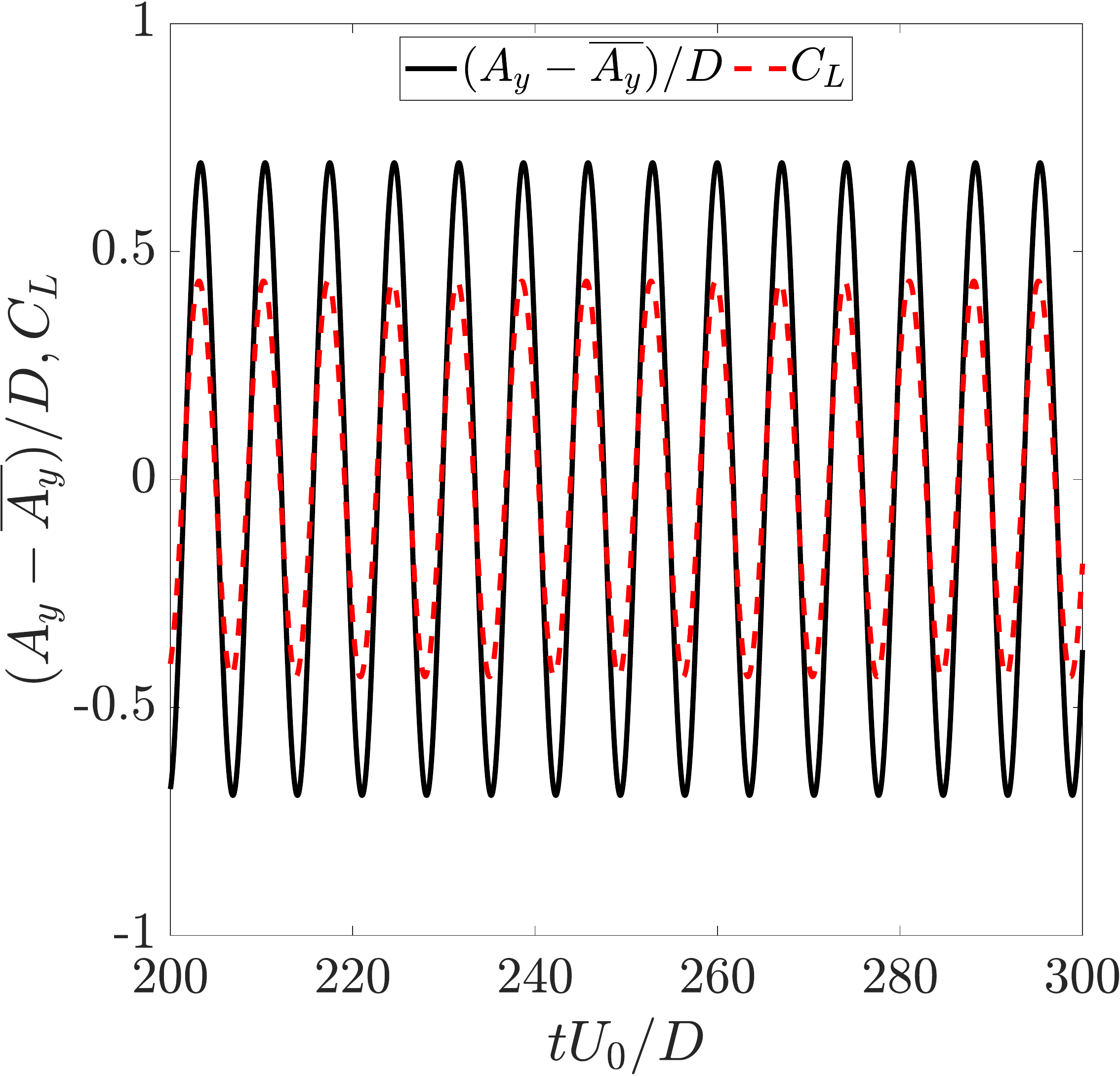}
\caption{}
\label{Ay-Cl}
\end{subfigure}%
\begin{subfigure}{0.5\textwidth}
\centering
\includegraphics[width=0.82\linewidth]{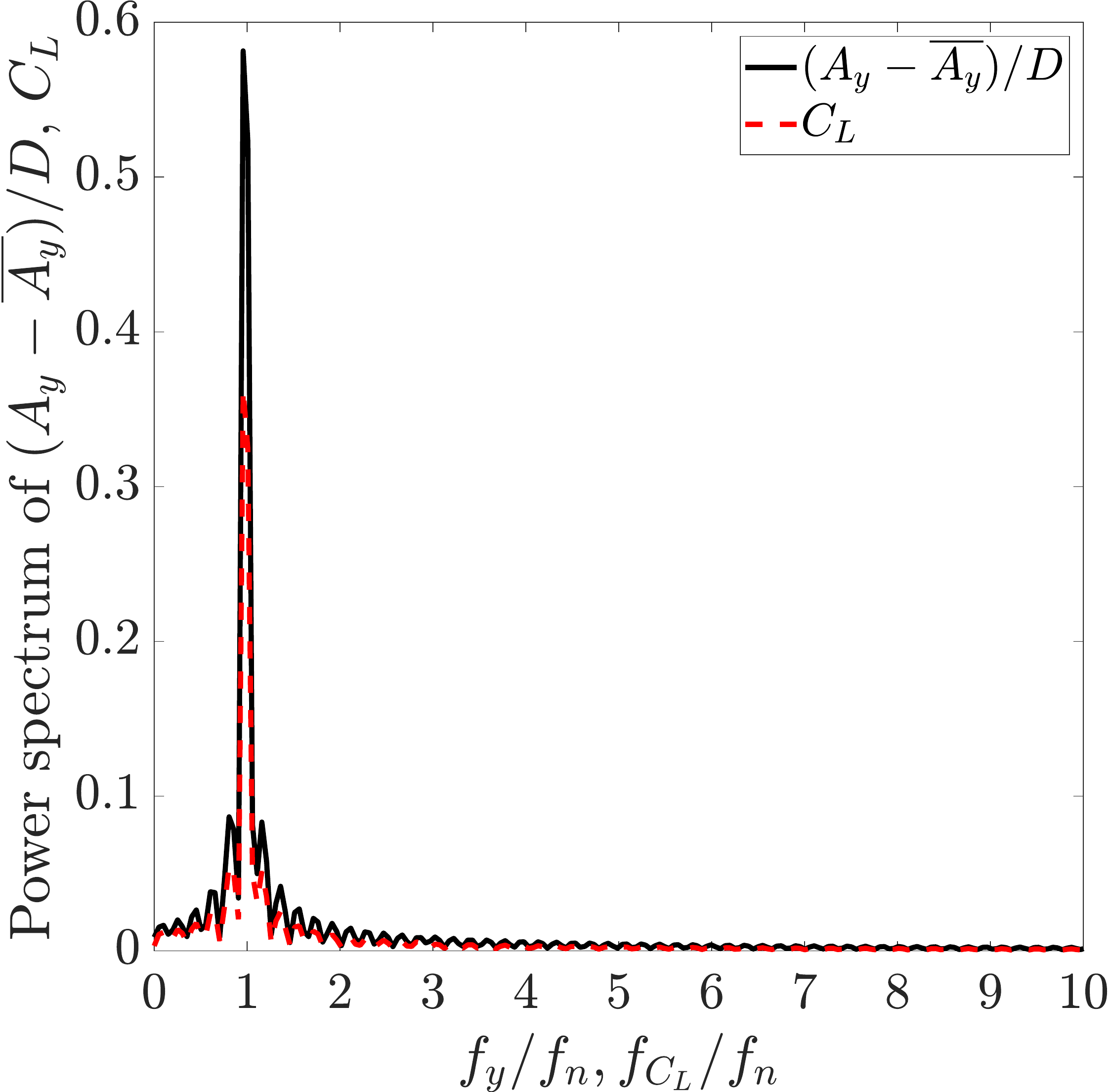}
\caption{}
\label{FFT-CLAy}
\end{subfigure}
\caption{(a) Variations of the dimensionless transverse vibration amplitude calculated from the mean deformed position of the cylinder $(A_\mathrm{y}-\overline{A_\mathrm{y}})/{D}$, probed at $z/L=1$, and lift coefficient $C_\mathrm{L}$ in time domain; (b) power spectra of the $(A_\mathrm{y}-\overline{A_\mathrm{y}})/{D}$ and $C_\mathrm{L}$ in frequency domain. The results are gathered in the time window $tU_{0}/D\in[200, 300]$ at $Re = 40$, $m^* = 1$, and $U^* = 7$.}
\label{CLAy-main}
\end{figure*}

Fig.~\ref{CLAy-main}\subref{Ay-Cl} demonstrates the time histories of the transverse vibration amplitude calculated from the mean deformed position of the cylinder $(A_\mathrm{y}-\overline{A_\mathrm{y}})/{D}$ at $z/L=1$ and lift coefficient $C_\mathrm{L}$ at $Re = 40$, $m^* = 1$, and $U^* = 7$. We observe that the transverse vibrations at the tip of the cylinder are in-phase with the variations of the lift coefficient. In addition, we show that the peak of the dimensionless transverse vibration frequency ($f_{\mathrm{y}}/f_\mathrm{n}$) matches the peak of the dimensionless lift coefficient frequency ($f_{\mathrm{C_L}}/f_\mathrm{n}$) in the frequency domain at $f_{\mathrm{y}}/f_\mathrm{n}=f_{\mathrm{C_L}}/f_\mathrm{n}=1$ (see Fig.~\ref{CLAy-main}\subref{FFT-CLAy}). This frequency match indicates that the lock-in phenomenon is driving the oscillations at $U^*=7$.

\begin{figure}[b]
\includegraphics[width=1\linewidth]{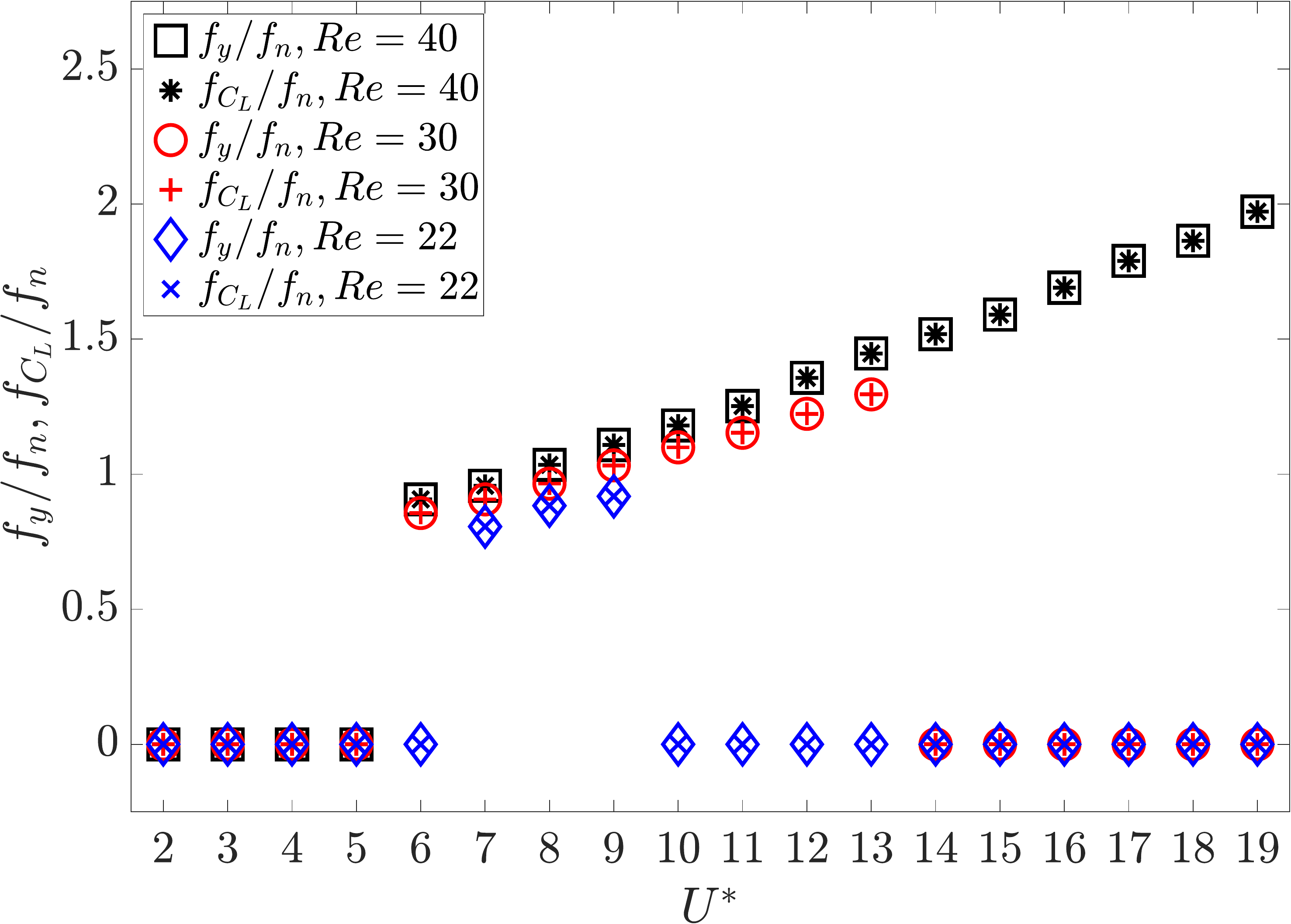}
\caption{\label{f_fn_Re40}Variations of the dimensionless transverse vibration frequency $f_\mathrm{y}/f_\mathrm{n}$, probed at $z/L=1$, and lift coefficient frequency $f_\mathrm{C_L}/f_\mathrm{n}$ with respect to $U^*$. The results are gathered at $m^*=1$ for $Re = 22,30$ and $40$.}
\end{figure}

To specify the range of the lock-in regime, we have provided the variations of the dimensionless transverse vibration frequency $f_{\mathrm{y}}/f_\mathrm{n}$ and lift coefficient frequency $f_{\mathrm{C_L}}/f_\mathrm{n}$ at $m^*=1$ for $Re=22, 30$ and $40$ with respect to $U^*$ in Fig.~\ref{f_fn_Re40}. We show that at $Re=22$, $f_{\mathrm{y}}/f_\mathrm{n}$ and $f_{\mathrm{C_L}}/f_\mathrm{n}$ are close to unity for $U^*\in[7,9]$ and zero elsewhere. However, for $Re=30$ and $40$, the lock-in regime is found to extend to a broader range of reduced velocities within $U^*\in[6,13]$. 

\begin{figure*}
    \begin{subfigure}{1\textwidth}
    \centering
    \includegraphics[width=0.55\linewidth]{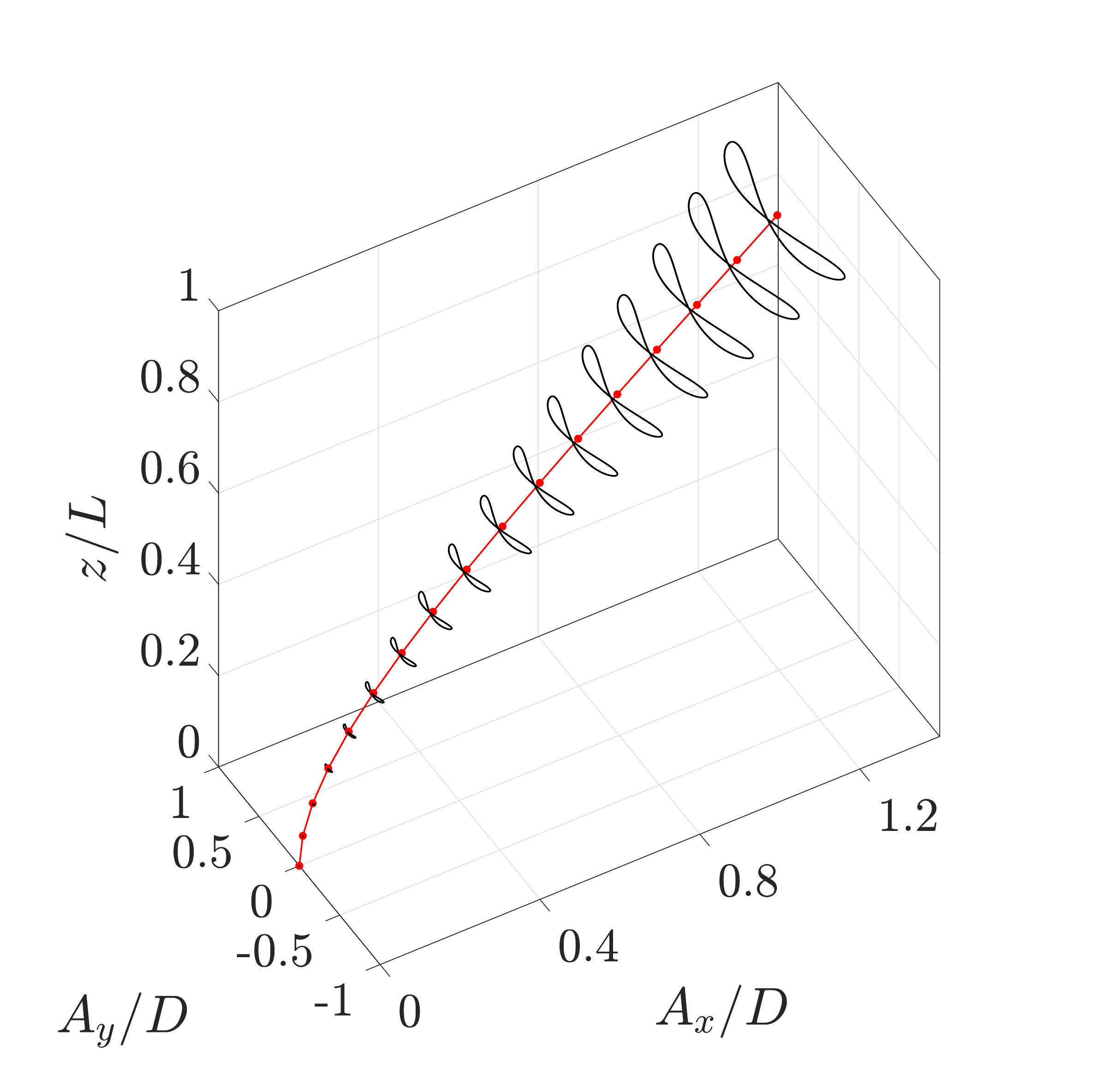}
    \caption{}
    \label{trajectory}
    \end{subfigure}
    \begin{subfigure}{0.5\textwidth}
    \centering
    \includegraphics[width=0.95\linewidth]{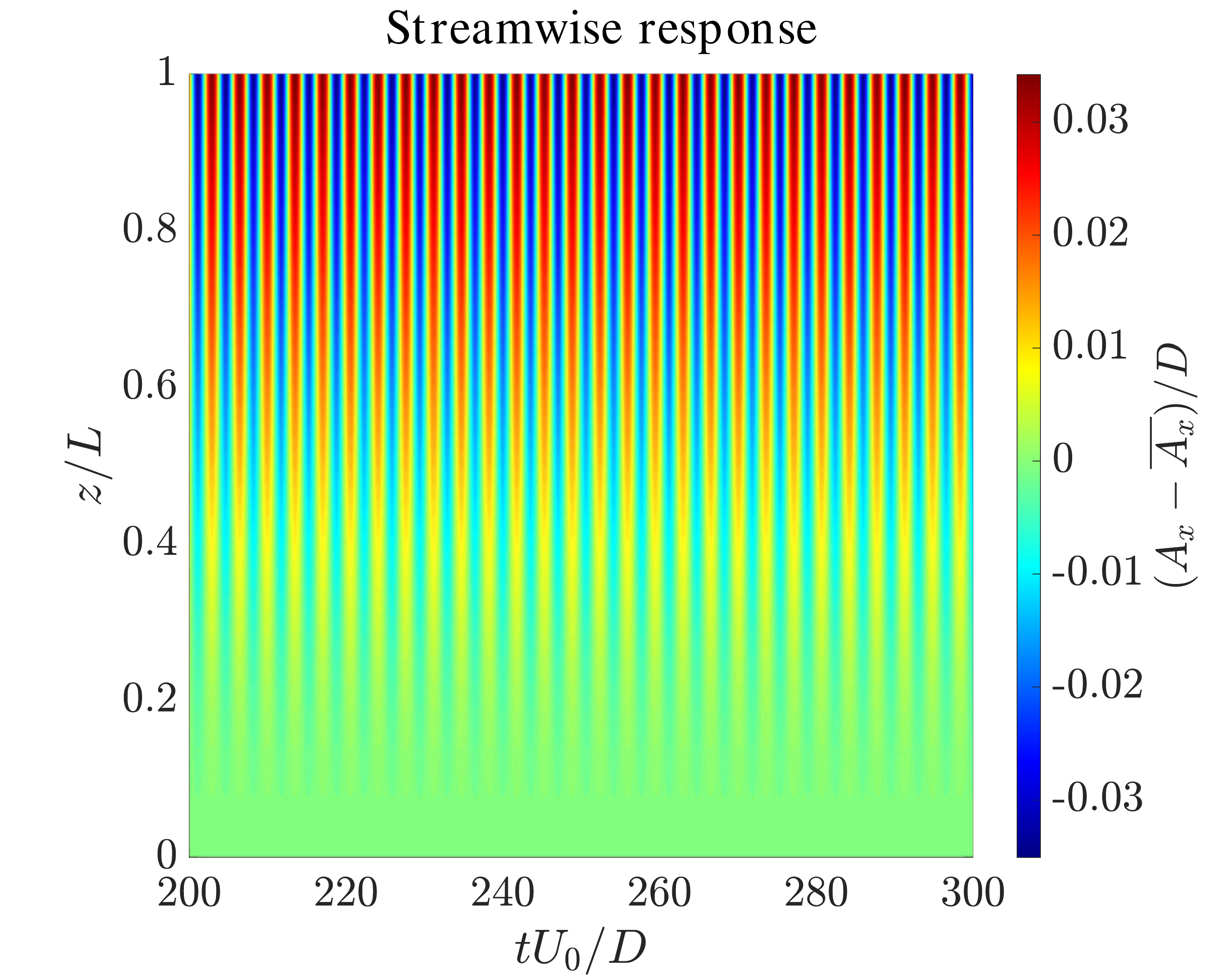}
    \caption{}
    \label{Dx-scalogram}
    \end{subfigure}%
    \begin{subfigure}{0.5\textwidth}
    \centering
    \includegraphics[width=1\linewidth]{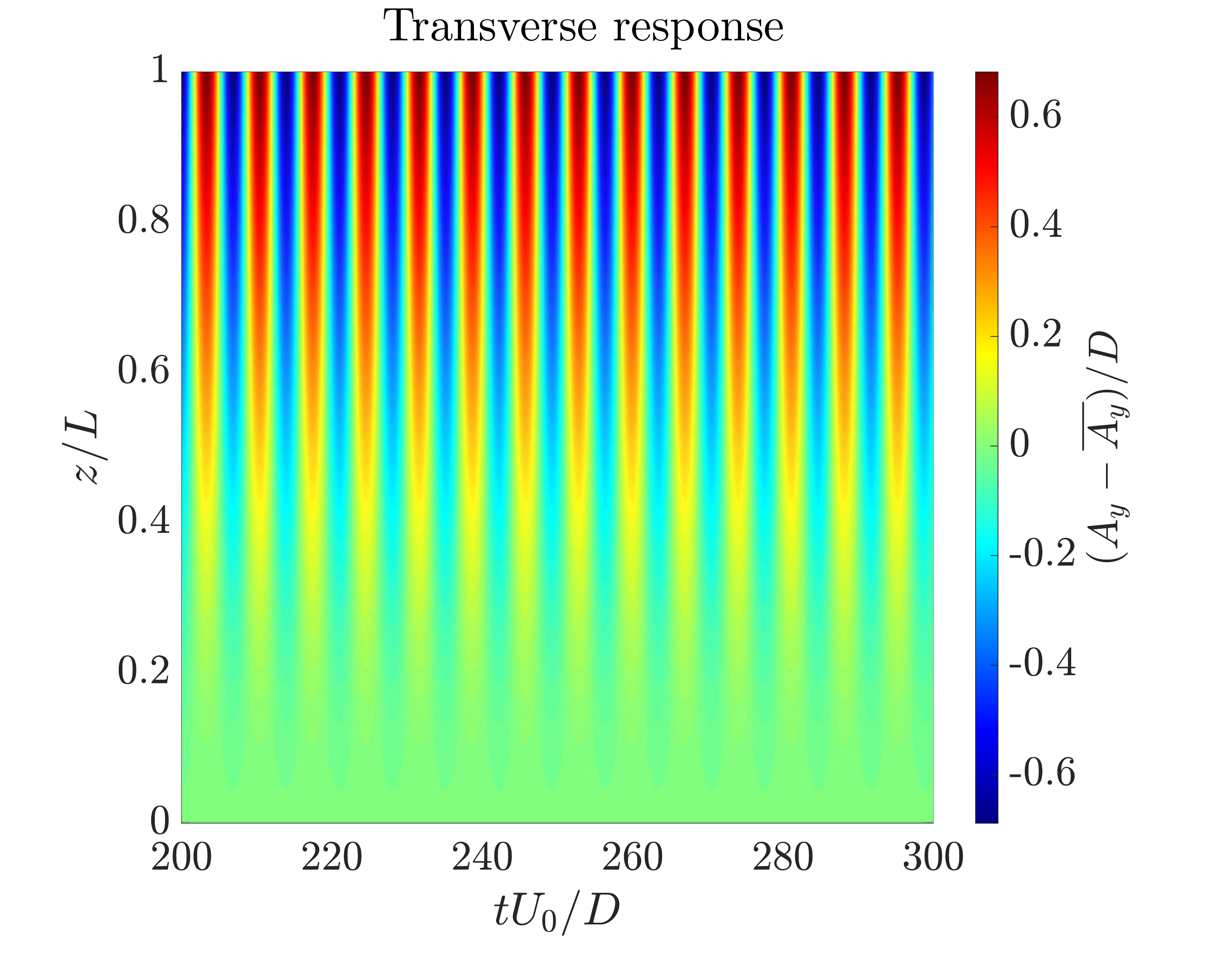}
    \caption{}
    \label{Dy-scalogram}
    \end{subfigure}
\caption{\label{trajectory-scalograms}(a) Motion trajectory of the flexible cantilever cylinder (illustrated by black lines); the red filled dots represent the mean position of the cylinder nodes and the red line corresponds to the cylinder's steady deflected position. (b) Scalogram of the vibrations in the streamwise direction. (c) Scalogram of the vibrations in the transverse direction. The results are gathered at $Re = 40$, $m^* = 1$, and $U^* = 7$ in the time window $tU_{0}/D\in[200, 300]$.}
\end{figure*}

An isometric view of the cylinder undergoing large-amplitude oscillations in the lock-in regime at $Re = 40$, $m^* = 1$, and $U^* = 7$ is illustrated in Fig.~\ref{trajectory-scalograms}~\subref{trajectory}. A figure-eight type motion trajectory is observed across the cylinder length. These trajectories are shown to grow in magnitude by moving towards the tip of the cylinder. The scalograms of the dynamic response of the cylinder in the streamwise and transverse directions are given in Figs.~\ref{trajectory-scalograms}~\subref{Dx-scalogram} and~\subref{Dy-scalogram}, respectively. We show that the cylinder exhibits a standing wave response, with oscillations being in the first mode of vibration in both the streamwise and transverse directions.

\begin{figure*}
\begin{subfigure}{0.4\textwidth}
\centering
\includegraphics[width=1\linewidth]{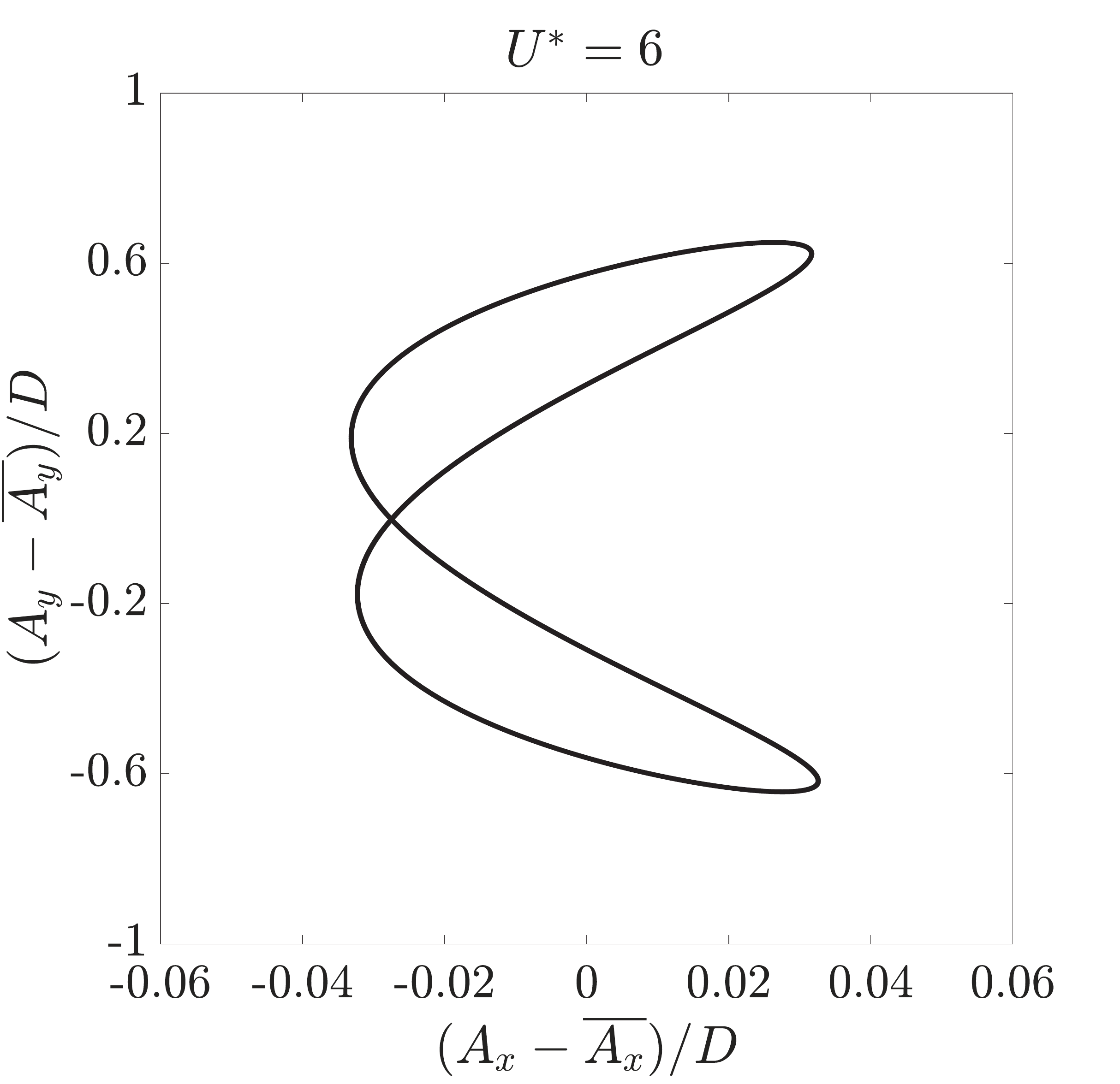}
\label{tipMotion-Ur4}
\end{subfigure}%
\begin{subfigure}{0.4\textwidth}
\centering
\includegraphics[width=1\linewidth]{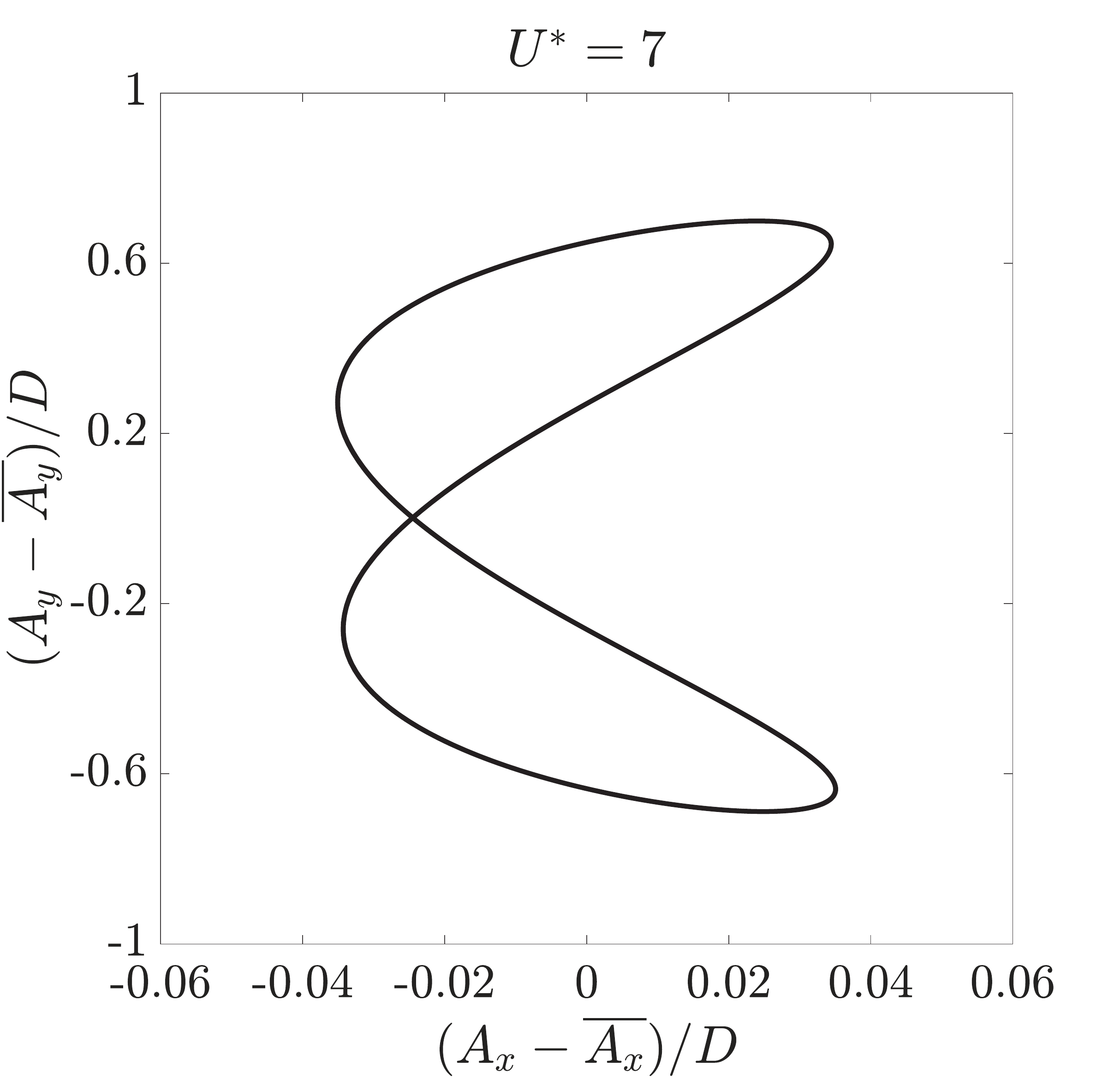}
\label{tipMotion-Ur5}
\end{subfigure}
\begin{subfigure}{0.4\textwidth}
\centering
\includegraphics[width=1\linewidth]{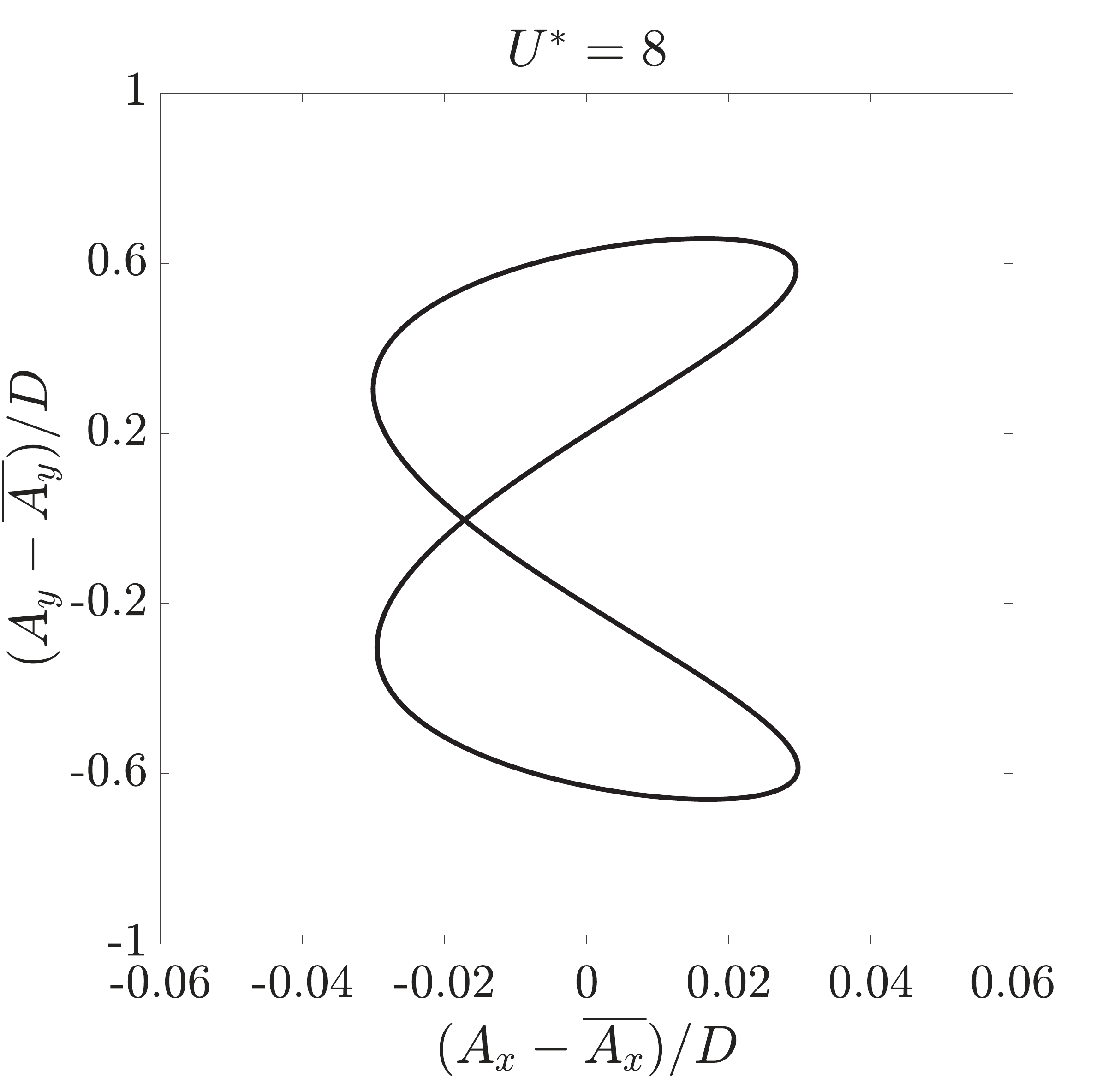}
\label{tipMotion-Ur6}
\end{subfigure}%
\begin{subfigure}{0.4\textwidth}
\centering
\includegraphics[width=1\linewidth]{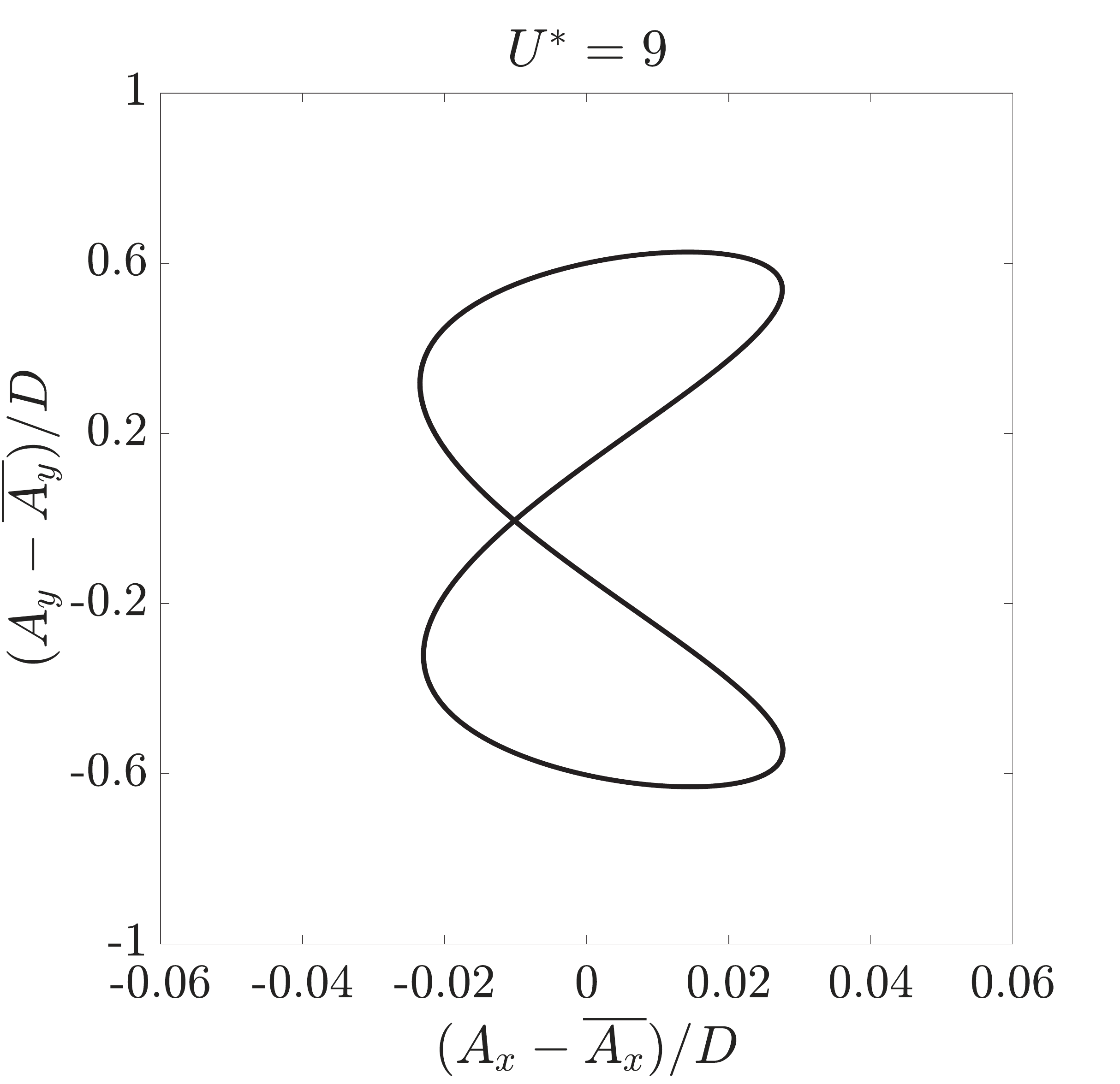}
\label{tipMotion-Ur7}
\end{subfigure}
\begin{subfigure}{0.4\textwidth}
\centering
\includegraphics[width=1\linewidth]{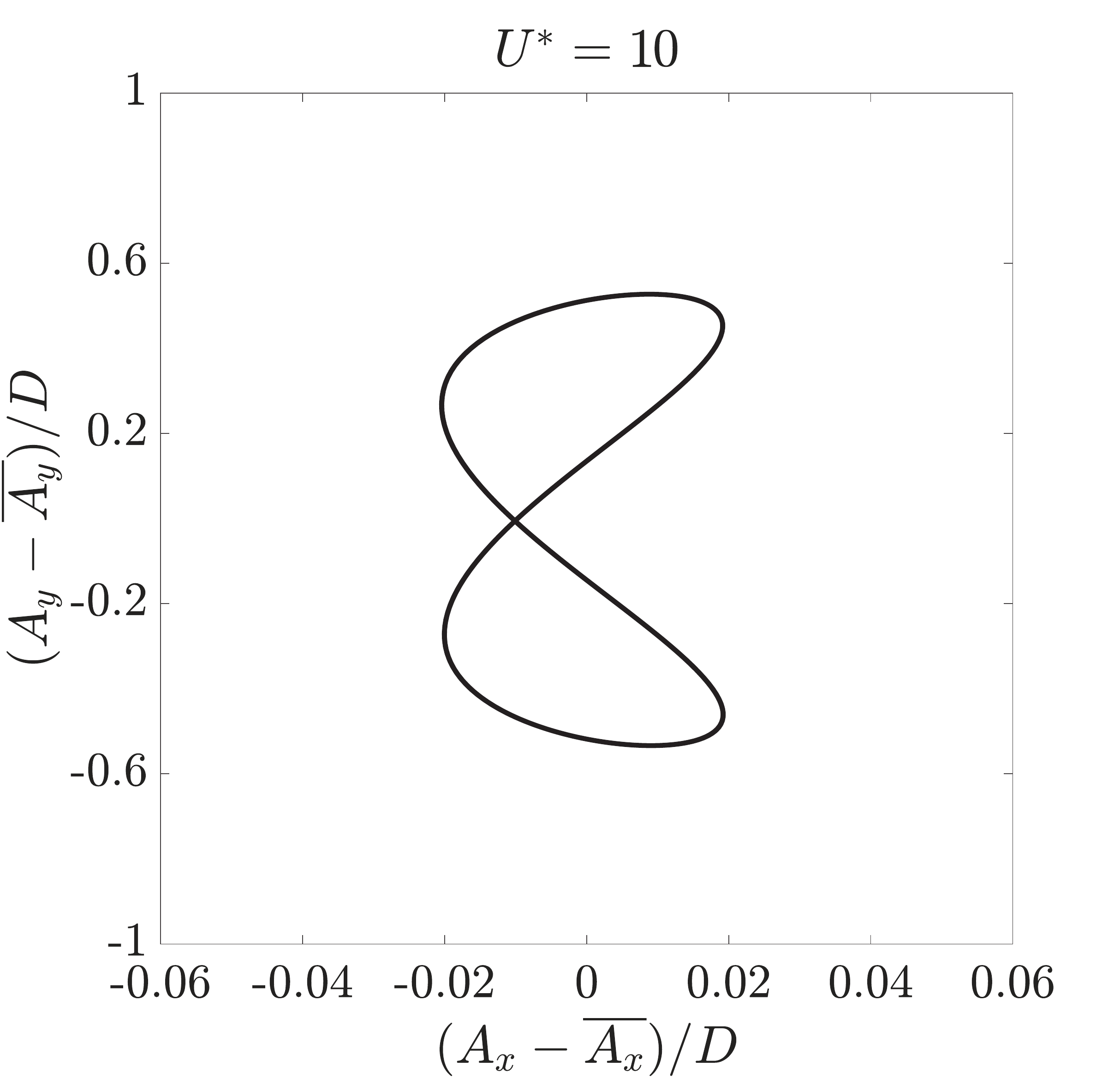}
\label{tipMotion-Ur8}
\end{subfigure}%
\begin{subfigure}{0.4\textwidth}
\centering
\includegraphics[width=1\linewidth]{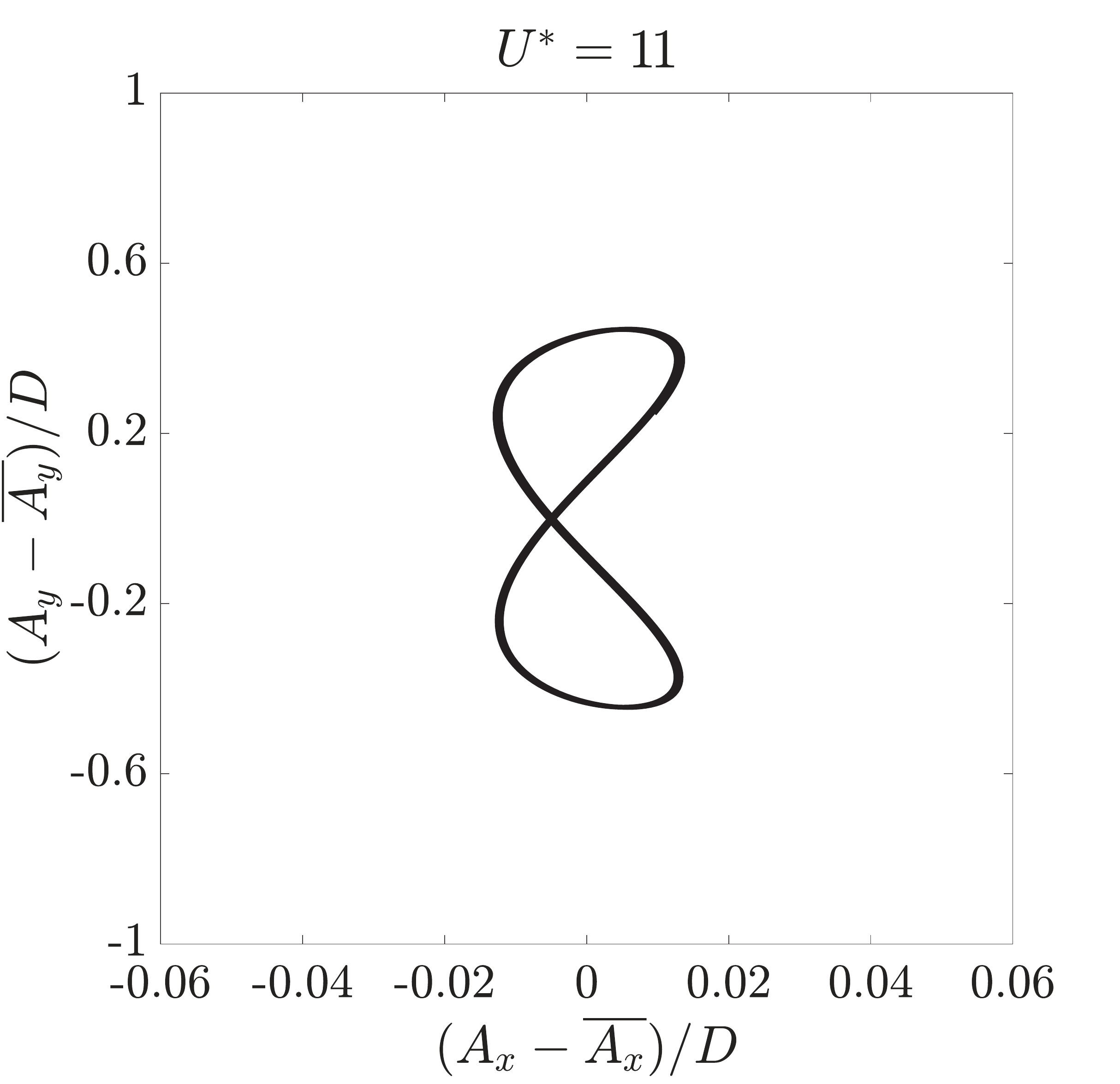}
\label{tipMotion-Ur9}
\end{subfigure}
\caption{\label{tipMotion}Motion trajectory of the flexible cantilever cylinder at $z/L=1$, $Re=40$, and $m^*=1$ for $U^*\in [6, 11]$.}
\end{figure*}

Based on the scalograms of the cylinder response, the dimensionless frequency of the streamwise oscillations ($f_{\mathrm{x}}/f_\mathrm{n}$) is found to be twice that of the dimensionless frequency of the transverse vibrations ($f_{\mathrm{y}}/f_\mathrm{n}$), i.e., $f_{\mathrm{x}}/f_\mathrm{n}=2f_{\mathrm{y}}/f_\mathrm{n}~\approx2$. This type of oscillatory response has been previously observed in two-degrees-of-freedom elastically mounted rigid cylinders undergoing VIVs~\cite{Li2016}. 
The motion trajectory of the tip of the cylinder at $Re = 40$, and $m^* = 1$ is given in Fig.~\ref{tipMotion} for $U^*\in [6, 11]$. The figure-eight shape of the motion trajectories is associated with the frequency ratio of $f_{\mathrm{x}}/f_\mathrm{y}\approx2$ in the lock-in regime. 
\subsection{Wake dynamics during lock-in}
\begin{figure*}
\begin{subfigure}{1\textwidth}
\centering
\includegraphics[width=0.35\linewidth]{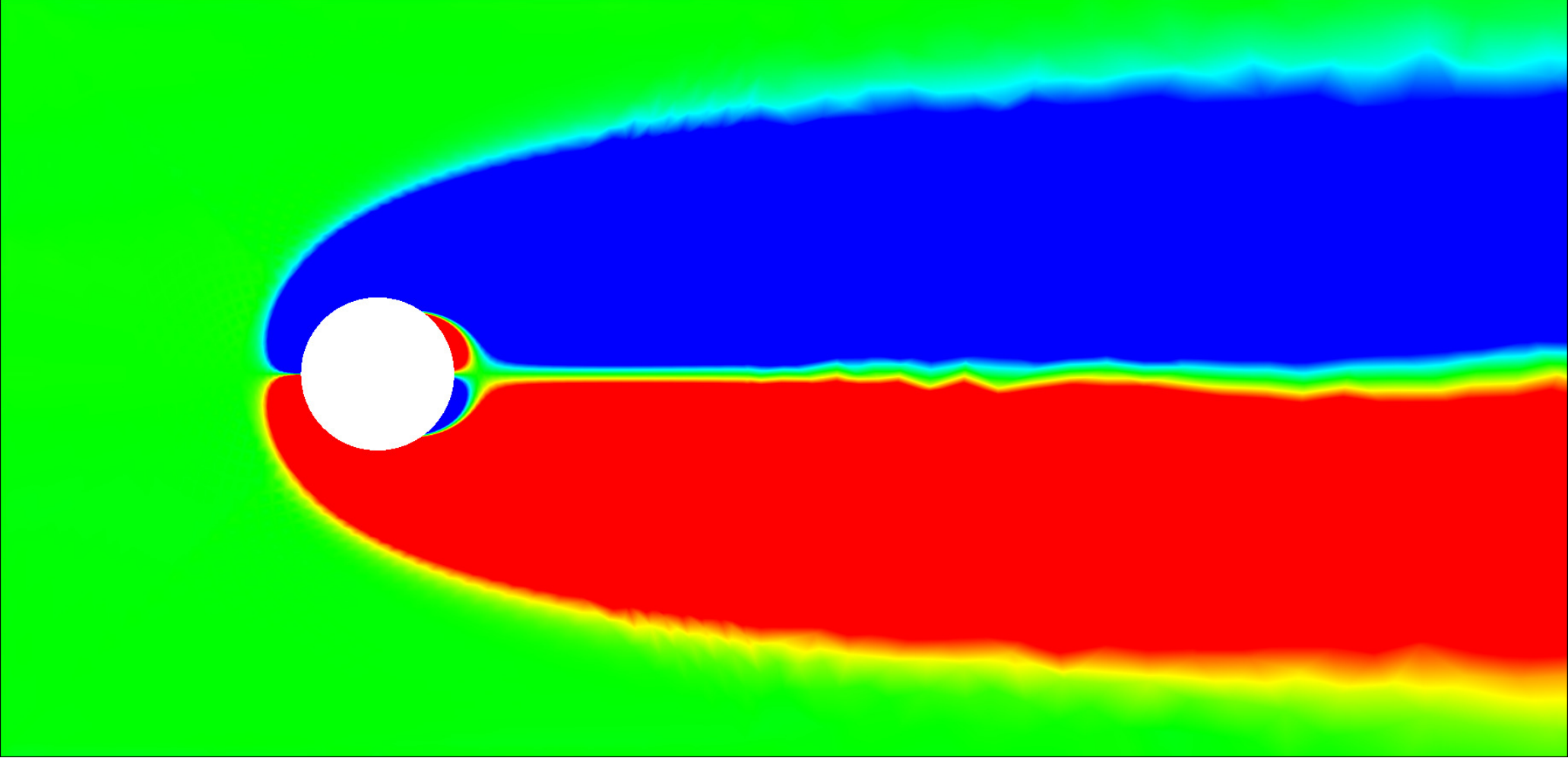}
\caption{}
\label{Re40-Ur5-t1point95}
\end{subfigure}
\begin{subfigure}{1\textwidth}
\centering
\includegraphics[width=0.35\linewidth]{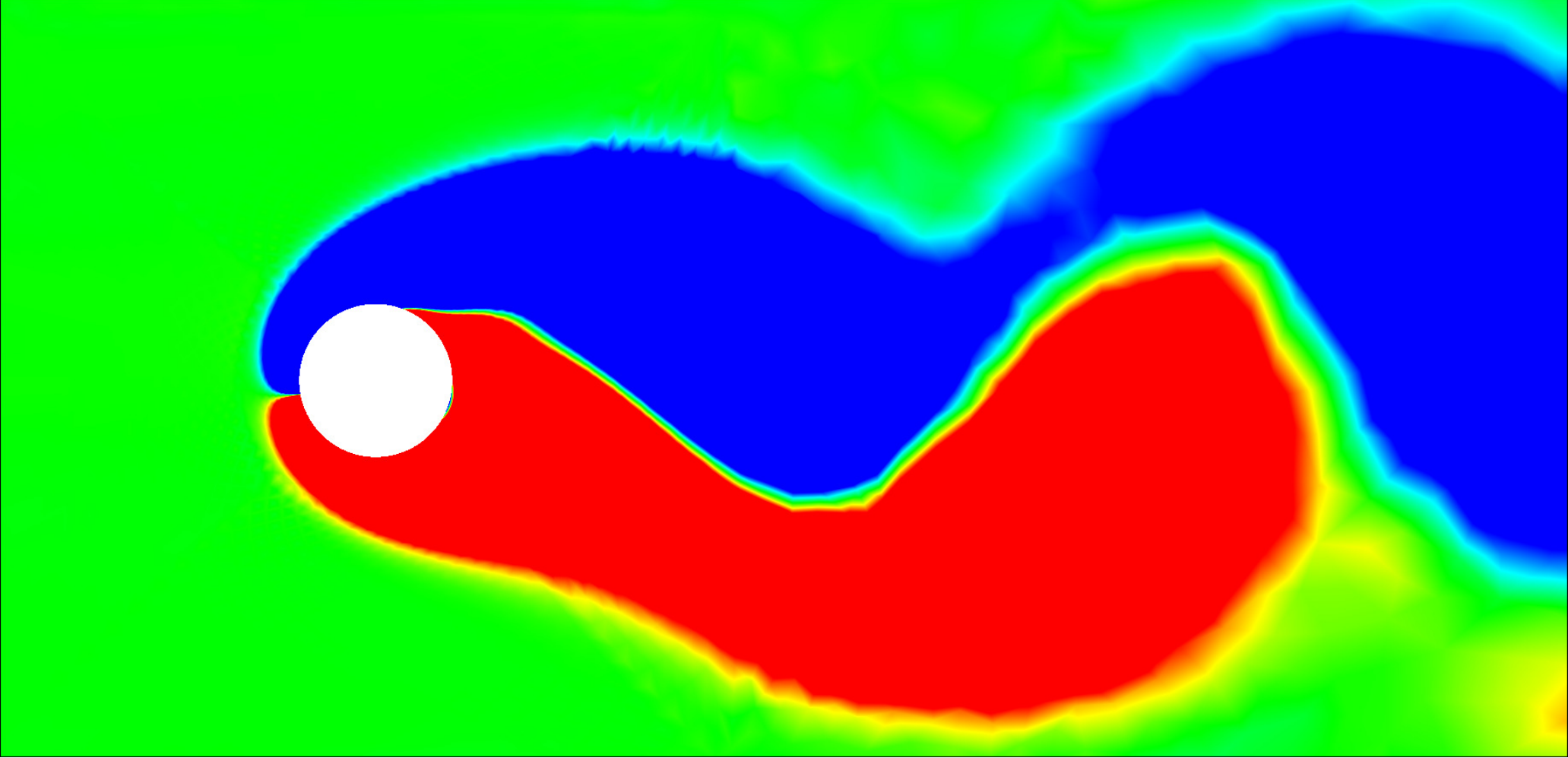}
\caption{}
\label{t3point73z1over2}
\end{subfigure}
\begin{subfigure}{1\textwidth}
\centering
\includegraphics[width=0.4\linewidth]{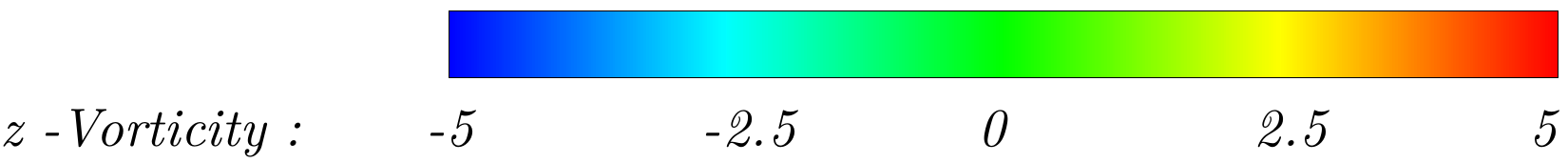}
\end{subfigure}
\caption{\label{Re40-Ur7-zVor-comparison}Comparison of the z-vorticity contour for the (a) stationary rigid cylinder and (b) flexible cantilever cylinder at $z/L=0.5$, $Re=40$, $m^*=1$, $U^* = 7$, and $tU_{0}/D=200$.}
\end{figure*}

\begin{figure*}
\begin{subfigure}{0.25\textwidth}
\centering
\includegraphics[width=0.75\linewidth]{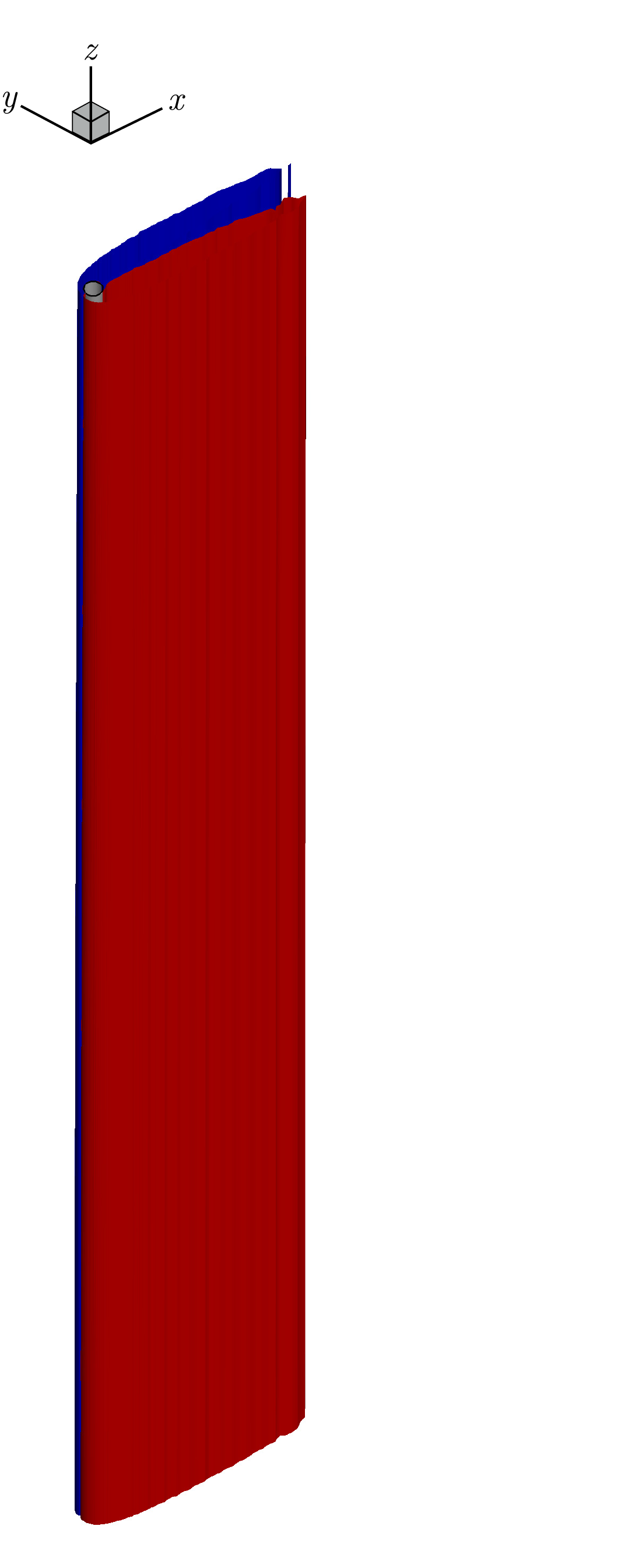}
\caption{$U^*=3$}
\label{isosurface-Re30-Ustar3}
\end{subfigure}%
\begin{subfigure}{0.25\textwidth}
\centering
\includegraphics[width=0.75\linewidth]{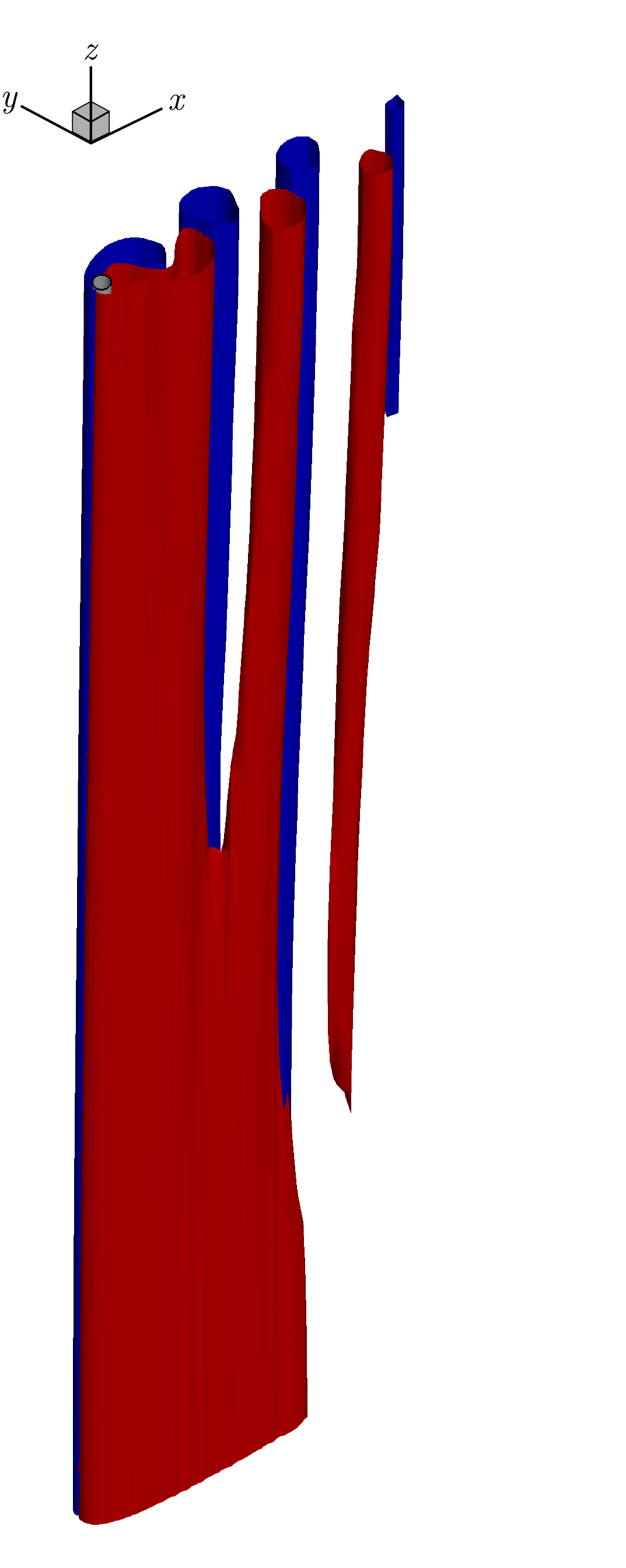}
\caption{$U^*=6$}
\label{isosurface-Re30-Ustar6}
\end{subfigure}%
\begin{subfigure}{0.25\textwidth}
\centering
\includegraphics[width=0.75\linewidth]{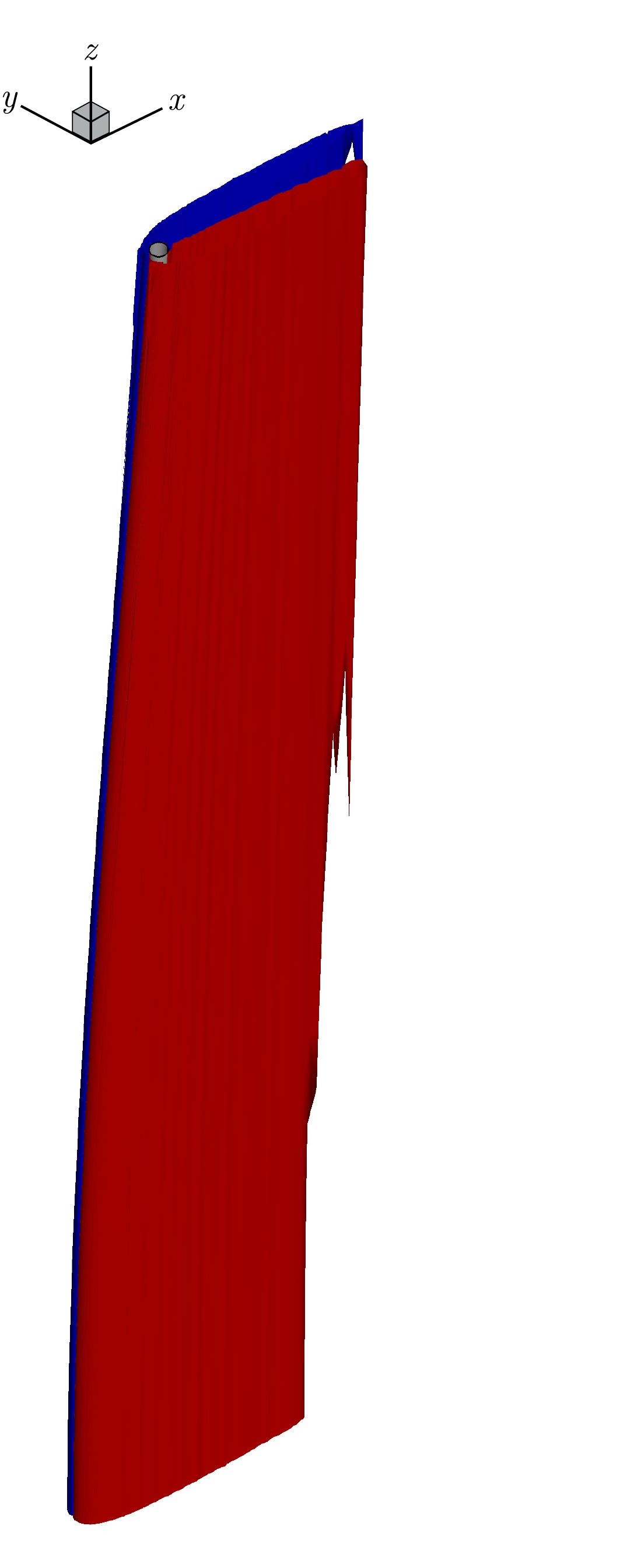}
\caption{$U^*=15$}
\label{isosurface-Re30-Ustar15}
\end{subfigure}
\caption{\label{zvort-isosurfaces-Re30-mstar1}Wake structures visualized by the normalized z-vorticity iso-surfaces ($\omega_{z}D/U_{0} = -0.224,0.224$) for the flexible cantilever cylinder at $Re=30$, and $m^*=1$. Red [blue] indicates regions of positive [negative] vortices.}
\end{figure*}

\begin{figure}[b]
\includegraphics[width=0.98\linewidth]{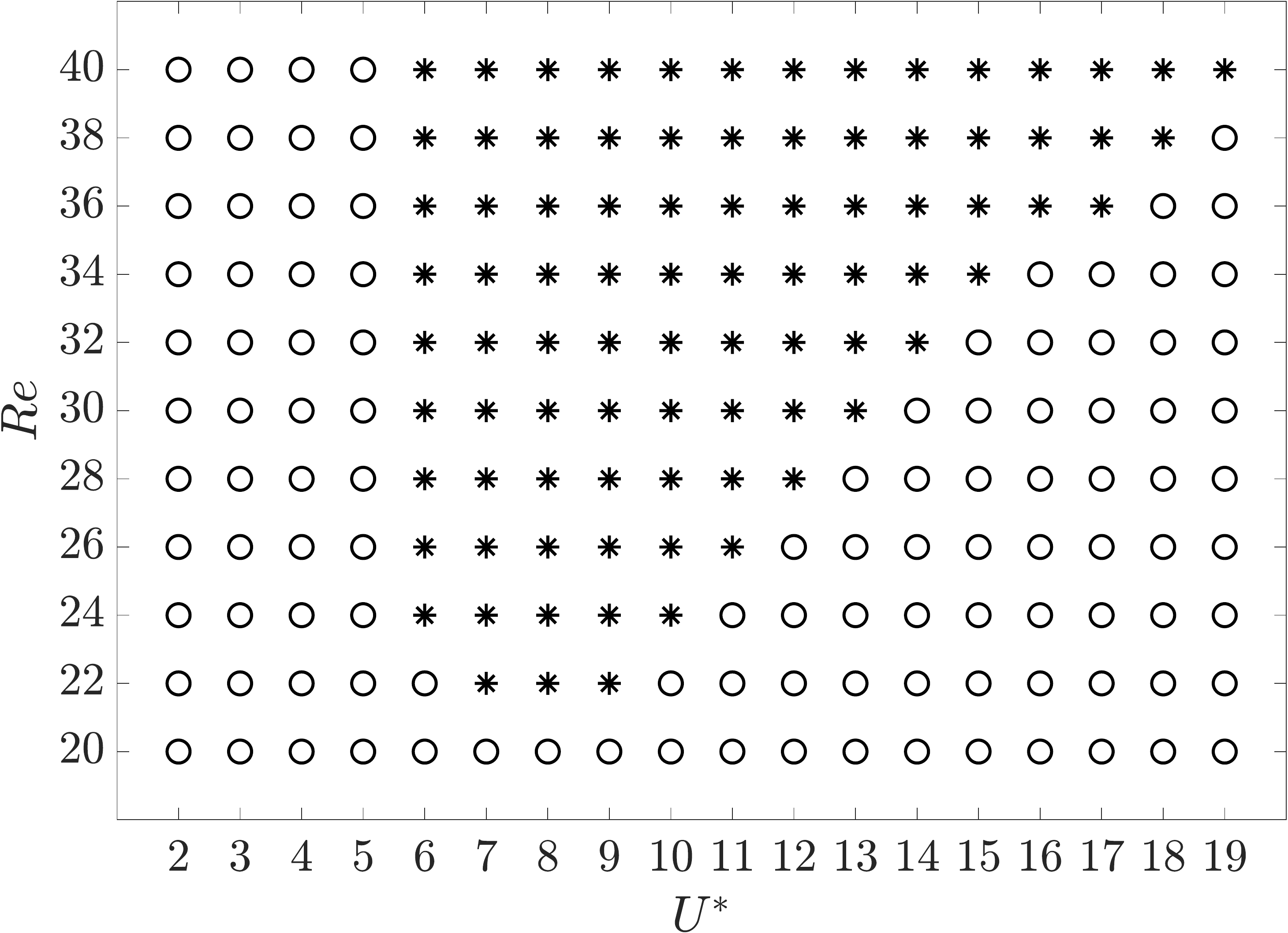}
\caption{\label{PhaseDiagram}Phase diagram of the wake stability as a function of $Re$ and $U^*$ at $m^* = 1$. Here, $\circ$ denotes a steady wake, while $*$ represents an unsteady wake behind the flexible cantilever cylinder.}
\end{figure}

Here, we examine the wake dynamics in the lock-in regime for the flexible cantilever cylinder at laminar subcritical $Re$. A comparison between the wake of a stationary rigid cylinder at $Re=40$, and the wake of the flexible cantilever cylinder at $z/L=0.5$, $Re=40$, $m^*=1$, and $U^* = 7$ is given in Fig.~\ref{Re40-Ur7-zVor-comparison}. 
We show that the wake of the stationary rigid cylinder is steady and symmetric with respect to the wake centerline at $Re = 40$; however, for the flexible cantilever cylinder, the wake is unstable at the same $Re$. To illustrate, we have examined the z-vorticity ($\omega_\mathrm{z}$) contours at different cross-sections of the flexible cantilever cylinder.
\begin{figure*}
\includegraphics[width=0.52\linewidth]{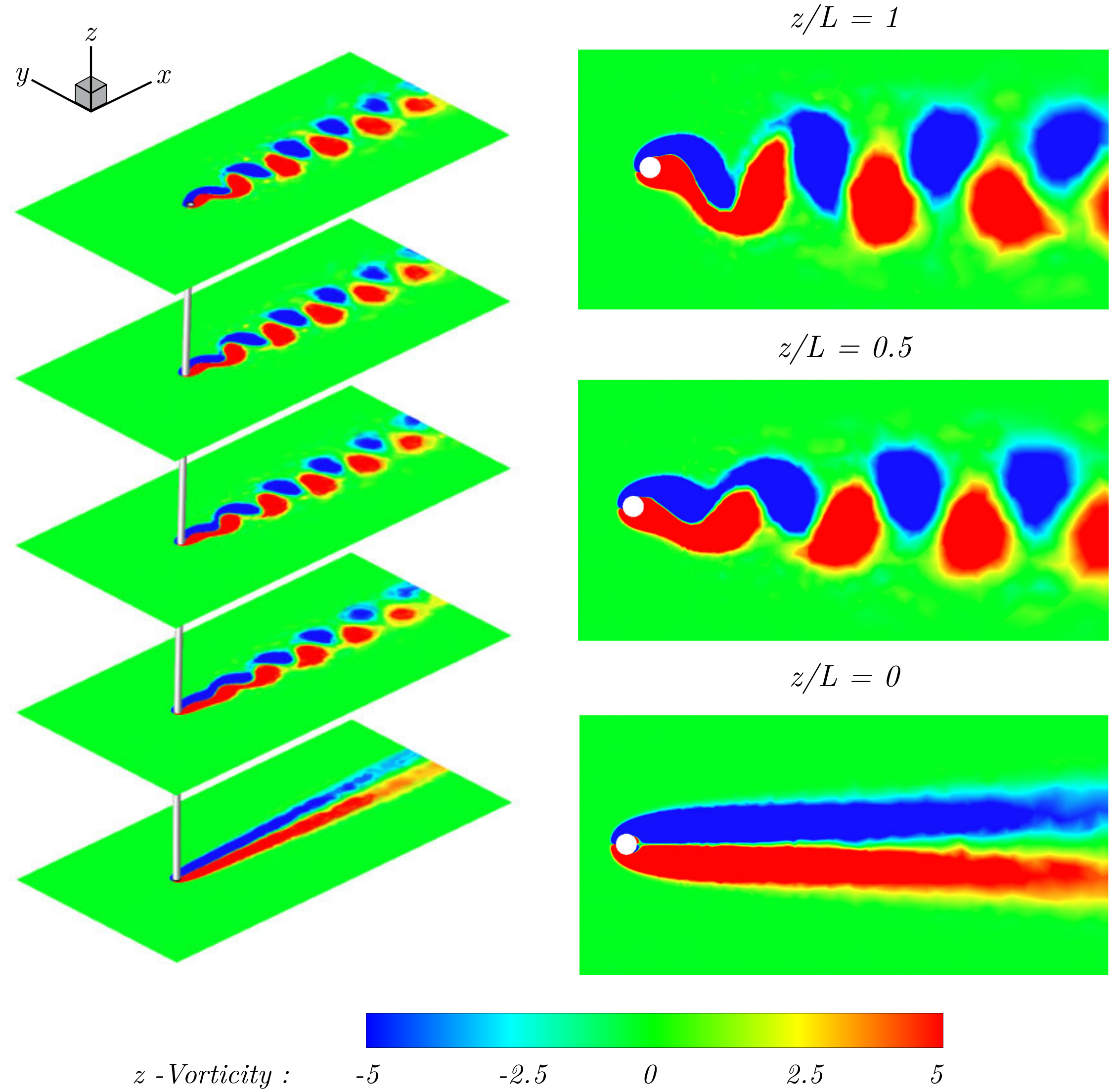}
\caption{\label{Z-Vorticity-New}Isometric view of the spanwise z-vorticity contours at various cross-sections of the flexible cantilever cylinder, with z-plane slices of the contours shown in the right-hand side for $z/L=1$, $0.5$, and $0$.}
\end{figure*}
As shown in Fig.~\ref{Z-Vorticity-New}, the wake of the cylinder is steady at $z/L=0$ where it is connected to fixed support; however, by approaching the tip of the cylinder, the flow starts to become unstable, and periodic vortex-shedding patterns are observed downstream. This finding suggests a connection between the cylinder motion and wake stability at laminar subcritical $Re$. To examine this conjecture, we have provided the z-vorticity iso-surfaces of the three-dimensional wake structures at $Re=30$ and $m^*=1$ in Fig.~\ref{zvort-isosurfaces-Re30-mstar1} for $U^*=3,6$ and $15$.
At the given $m^*$ and $Re$, $U^*=3,6$ and $15$ represent the pre-lock-in, lock-in, and post-lock-in regimes, respectively. We find that the flow field in the wake of the flexible cantilever cylinder is steady at $U^*=3$ (pre lock-in) and $U^*=15$ (post lock-in); however, an unsteady wake is observed at $U^*=6$ (see Fig.~\ref{zvort-isosurfaces-Re30-mstar1}). The phase diagram of the wake stability as a function of $Re$ and $U^*$ at $m^*=1$ is given in Fig.~\ref{PhaseDiagram}. 
We find that at $Re=20$, the flow field is steady for all $U^*$ values; however, as $Re$ is increased, the wake becomes unsteady for a particular range of reduced velocities. As shown in Fig.~\ref{PhaseDiagram}, the flow field in the wake of the flexible cantilever cylinder is unstable at $Re=22$ for $U^*\in[7,9]$. The range of the wake unsteadiness is shown to become wider at higher $Re$. For example, this range is between $U^*\in [6,13]$ at $Re=30$ and increases to $U^*\in [6,19]$ at $Re=40$. An important point to note here is that there is a critical $U^*\in[6,7]$ that marks the initiation of the wake unsteadiness for $22\leq Re\leq40$. This critical $U^*$ also marks the lower bound of the lock-in regime, as shown in the results of Section~\ref{subsec:responseCharacteristics}. Thus, we can infer that the range of the wake unsteadiness is closely correlated with the range of the lock-in regime at laminar subcritical $Re$. 

\begin{figure*}
\begin{subfigure}{0.25\textwidth}
\centering
\includegraphics[width=0.7\linewidth]{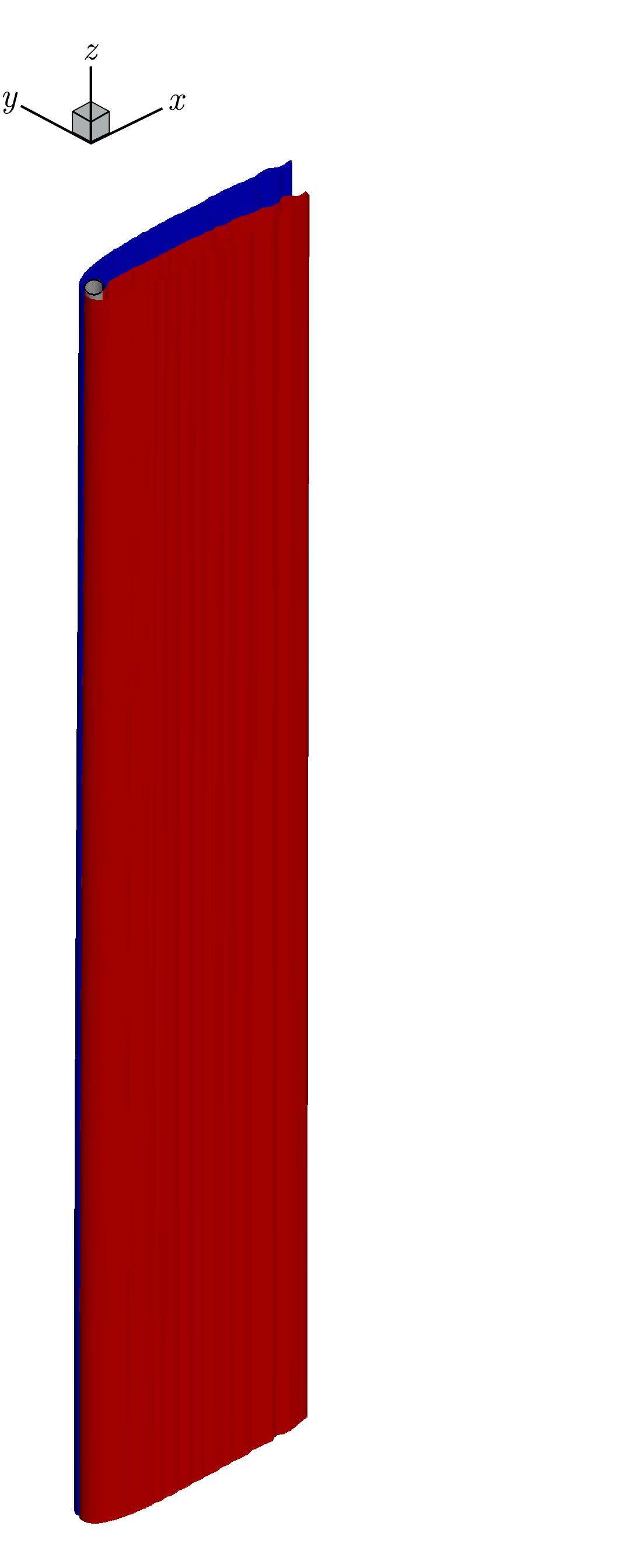}
\caption{$U^*=5$}
\label{isosurface-Ustar5}
\end{subfigure}%
\begin{subfigure}{0.25\textwidth}
\centering
\includegraphics[width=0.7\linewidth]{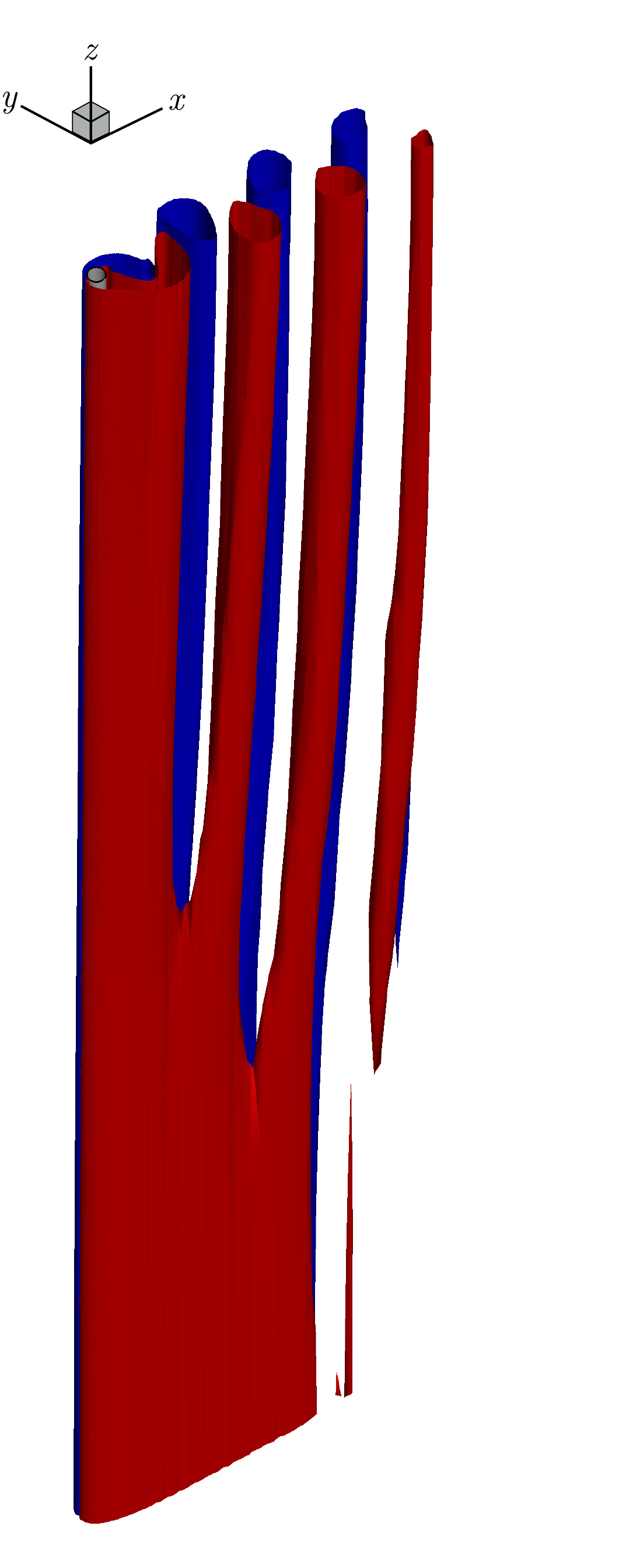}
\caption{$U^*=6$}
\label{isosurface-Ustar6}
\end{subfigure}%
\begin{subfigure}{0.25\textwidth}
\centering
\includegraphics[width=0.7\linewidth]{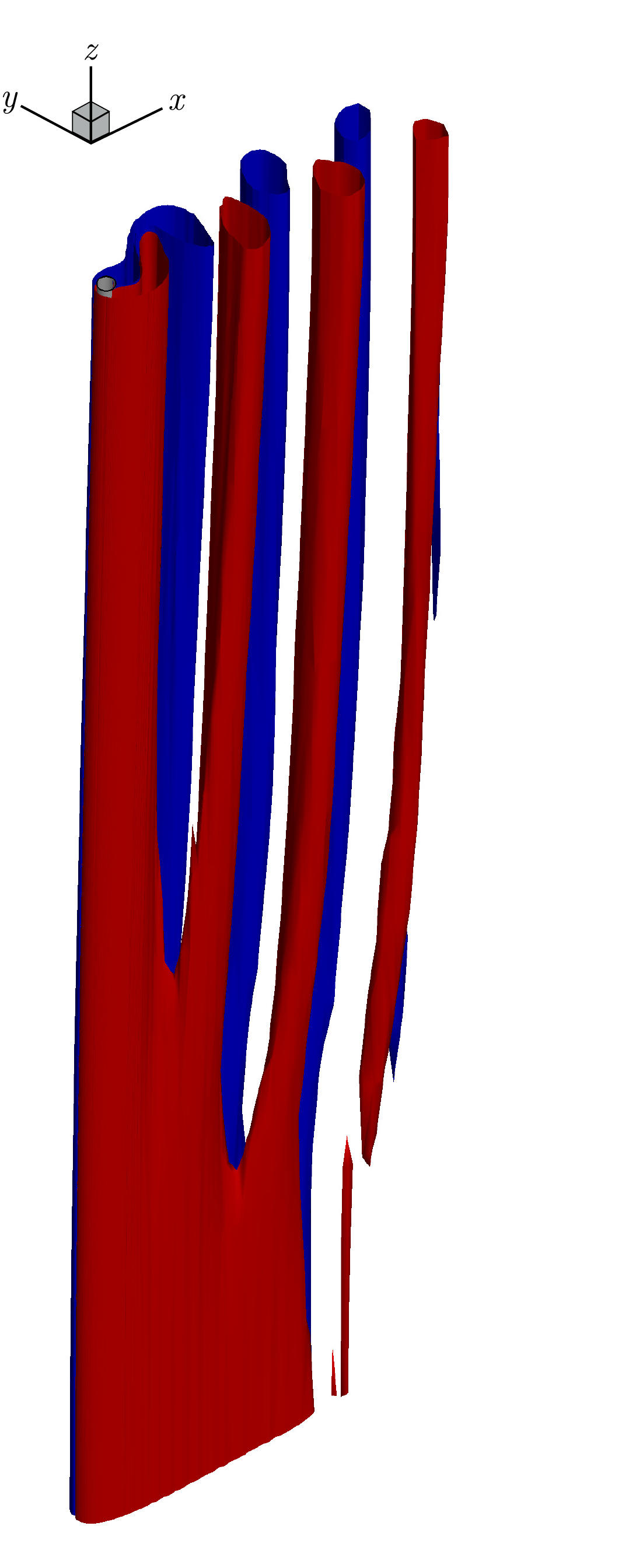}
\caption{$U^*=7$}
\label{isosurface-Ustar7}
\end{subfigure}%
\begin{subfigure}{0.25\textwidth}
\centering
\includegraphics[width=0.7\linewidth]{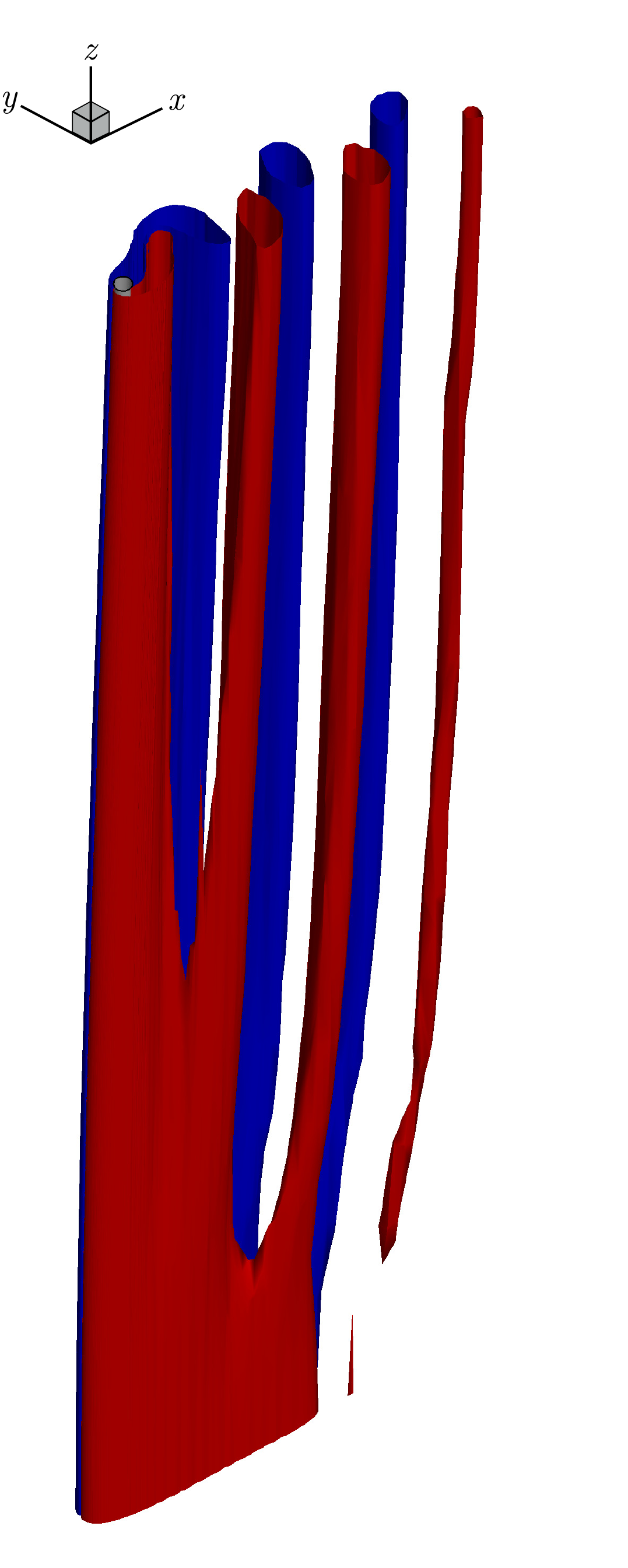}
\caption{$U^*=8$}
\label{isosurface-Ustar8}
\end{subfigure}
\begin{subfigure}{0.25\textwidth}
\centering
\includegraphics[width=0.7\linewidth]{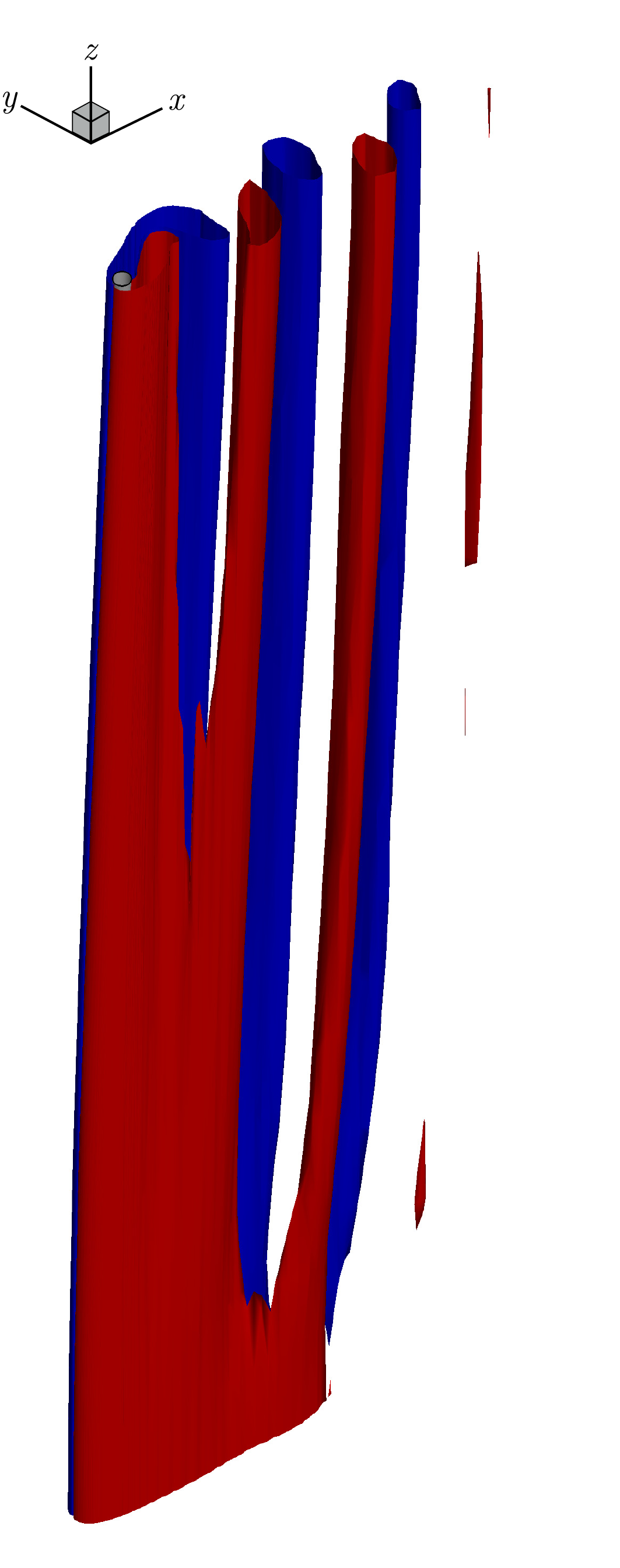}
\caption{$U^*=10$}
\label{isosurface-Ustar10}
\end{subfigure}%
\begin{subfigure}{0.25\textwidth}
\centering
\includegraphics[width=0.7\linewidth]{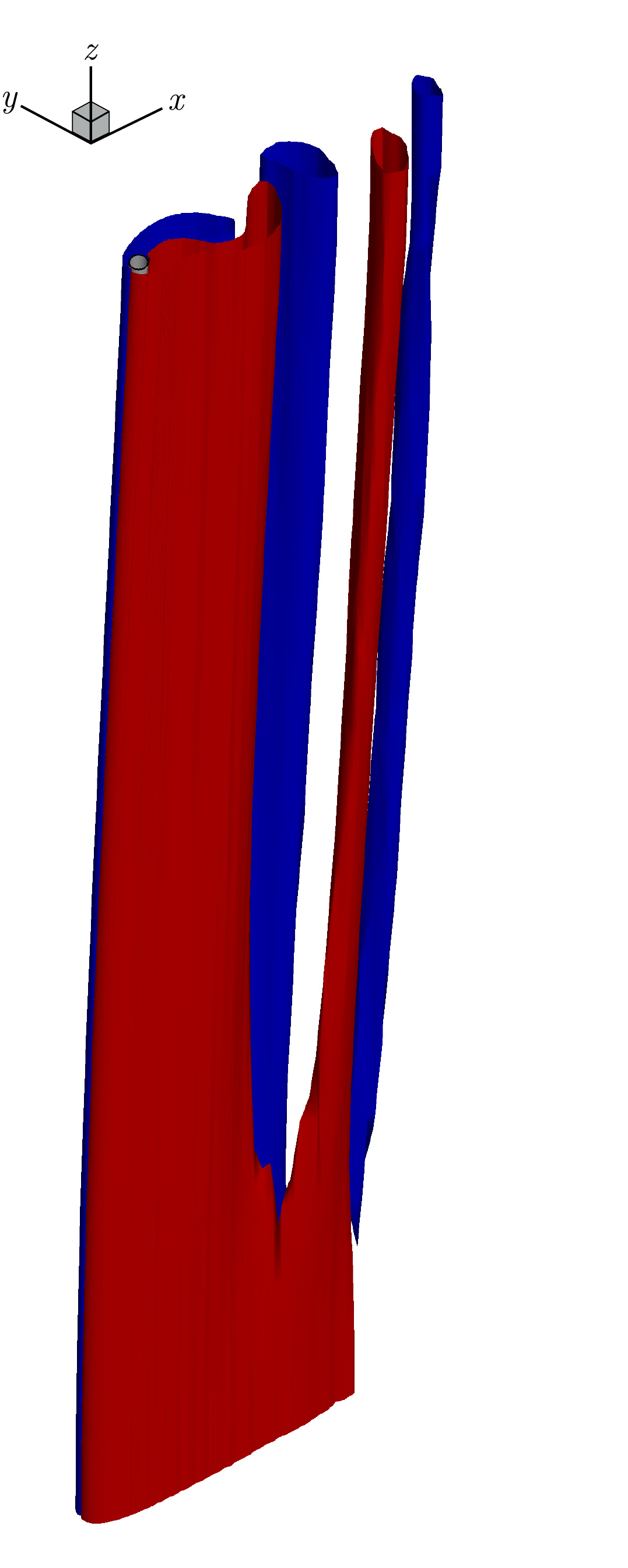}
\caption{$U^*=12$}
\label{isosurface-Ustar12}
\end{subfigure}%
\begin{subfigure}{0.25\textwidth}
\centering
\includegraphics[width=0.7\linewidth]{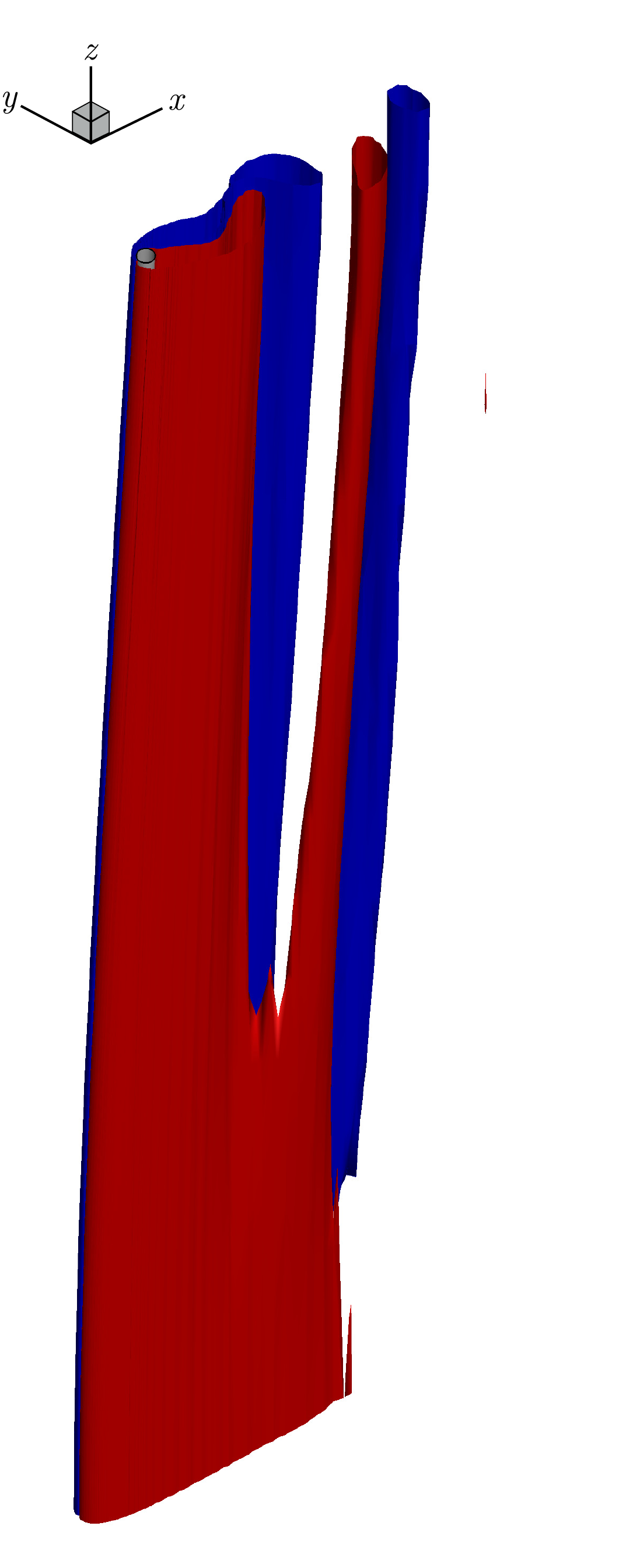}
\caption{$U^*=14$}
\label{isosurface-Ustar14}
\end{subfigure}
\caption{\label{zvort-isosurfaces}Wake structures visualized by the normalized z-vorticity iso-surfaces ($\omega_{z}D/U_{0} = -0.224,0.224$) for the flexible cantilever cylinder at $Re=40$, and $m^*=1$. Red [blue] indicates regions of positive [negative] vortices.}
\end{figure*}

We have provided the wake structures around the flexible cantilever cylinder at $Re=40$ and $m^*=1$ for $5\leq U^*\leq 14$ in Fig.~\ref{zvort-isosurfaces}.
At $U^*=5$, which represents the pre-lock-in regime, a steady wake flow is observed behind the cylinder; however, for $U^*\geq 6$, the wake is shown to become unstable, with two alternate vortices being shed from the cylinder wake in each cycle. In addition, the wake structures close to the fixed end of the cylinder are found to be steady, regardless of $U^*$. As shown in Fig.~\ref{zvort-isosurfaces}, although an unsteady wake is observed for $U^*\geq 6$ at distances between $z/L\in[0.2,1]$ from the fixed end of the cylinder, the wake is shown to be steady for $z/L\in[0,0.2]$. Thus, we can deduce that the three-dimensional flow phenomena do not contribute to the wake stability at this $Re$ regime. For the flow around an isolated cylinder, the wake has been shown to first become three-dimensional at $Re\approx200$~\cite{karniadakis_triantafyllou_1992}.

With these findings, we can deduce that the wake of a flexible cantilever cylinder could become unsteady at laminar subcritical $Re$, provided that two essential requirements are met: (i) the flow needs to have sufficiently large inertia to overcome the viscous damping and (ii) the system parameters need to be in the lock-in range to sustain the unsteadiness in the wake. 

In the next section, we discuss the relationship between the cylinder motion and stability of the wake in detail.
\subsection{Relationship between the cylinder dynamics and wake unsteadiness}
Here, we pinpoint the relationship between the cylinder motion and stability of the wake at laminar subcritical $Re$. We recommend a combined VIV-galloping type instability as the possible cause of the wake unsteadiness for $Re<Re_\mathrm{cr}$. Galloping is a velocity-dependent and damping-controlled fluid-structure instability, which is generally observed in geometrically asymmetric structures~\cite{Hartog1985}. Although the flow field around an asymmetric structure is uniform in magnitude and direction, cross-flow oscillations of the asymmetric body alter the magnitude and direction of the incident flow with respect to the body coordinate system. This change, in turn, alters the fluid forces acting on the body and could trigger the galloping instability. A deviation from symmetric cross-section in transmission lines due to ice formation~\cite{Farzaneh2008} or in marine cables due to marine organisms~\cite{Simpson1972} are some examples of the galloping instability in engineering structures. Galloping is known to cause large-amplitude sustained oscillations in flexible or elastically-mounted structures~\cite{Hartog1985}.

In contrast to vortex-induced vibrations, galloping instability is induced by a relative body motion rather than the unsteady fluctuations of the flow field; hence it can occur even for steady attached flows. When the transverse force acting on a flexible or elastically-mounted body increases in the direction of motion, it adds movement to the body, and the body will displace further until the opposing stiffness or damping overcomes the movements, or the transverse force decreases when the movement is increased. For a flexible cantilever cylinder interacting with fluid flow, the body is free to deform in the streamwise and transverse directions. Although displacements in the streamwise direction do not contribute to the stability of the wake~\cite{Tang1997,Vasconcelos2011}, relative movements in the transverse direction break the wake symmetry, altering the fluid forces acting on the cylinder. This symmetry breakdown, in turn, induces a galloping-type instability by creating negative damping in the combined fluid-structure system. The low-speed galloping-type instability, together with the frequency lock-in, is most arguably the mechanism that leads to sustained unsteadiness in the wake at laminar subcritical $Re$. 

\begin{figure}
\begin{subfigure}{0.22\textwidth}
\centering
\includegraphics[width=0.9\linewidth]{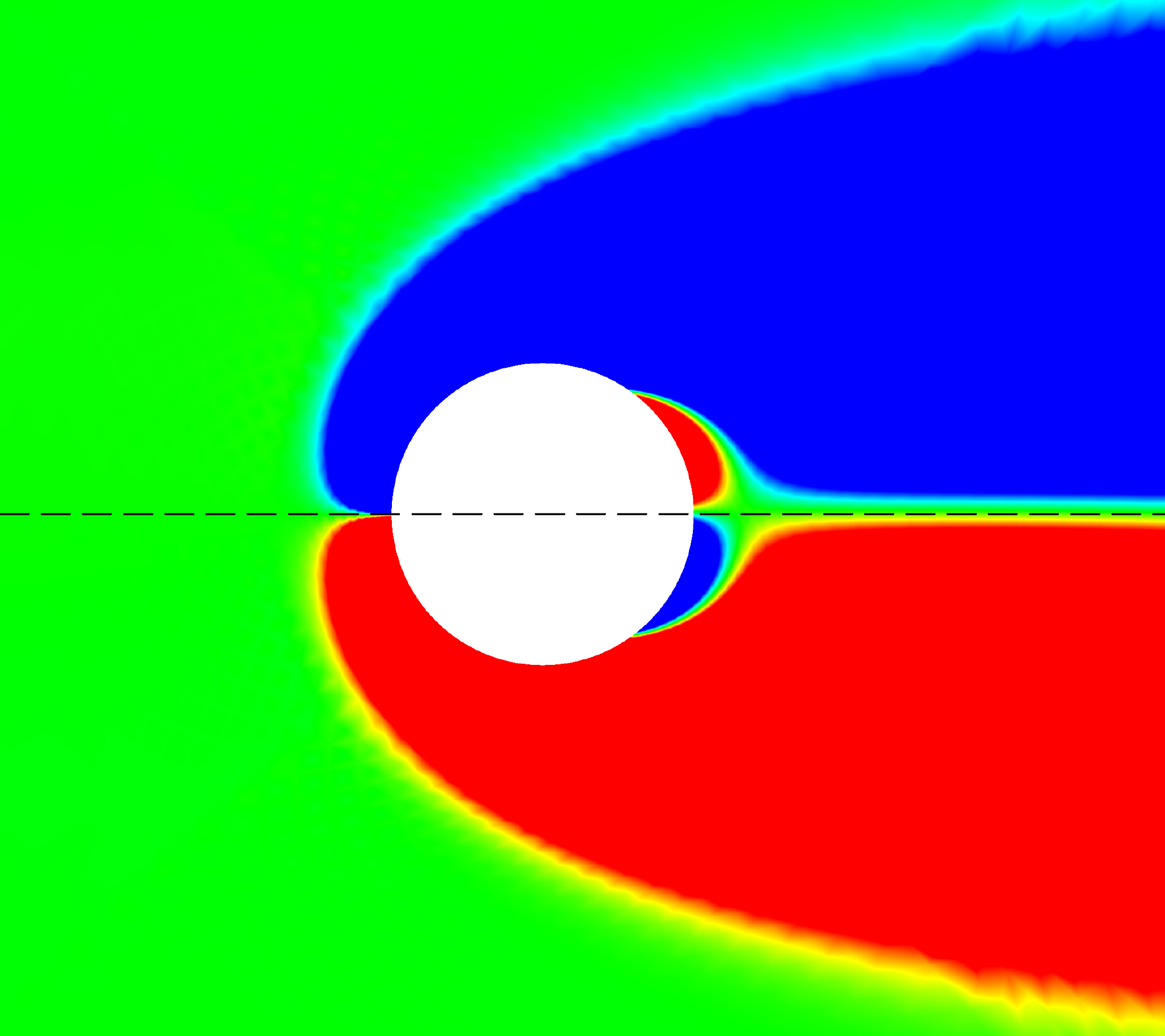}
\caption{$tU_{0}/D=60$}
\label{Re40-Ur5-t1point75}
\end{subfigure}%
\begin{subfigure}{0.22\textwidth}
\centering
\includegraphics[width=0.9\linewidth]{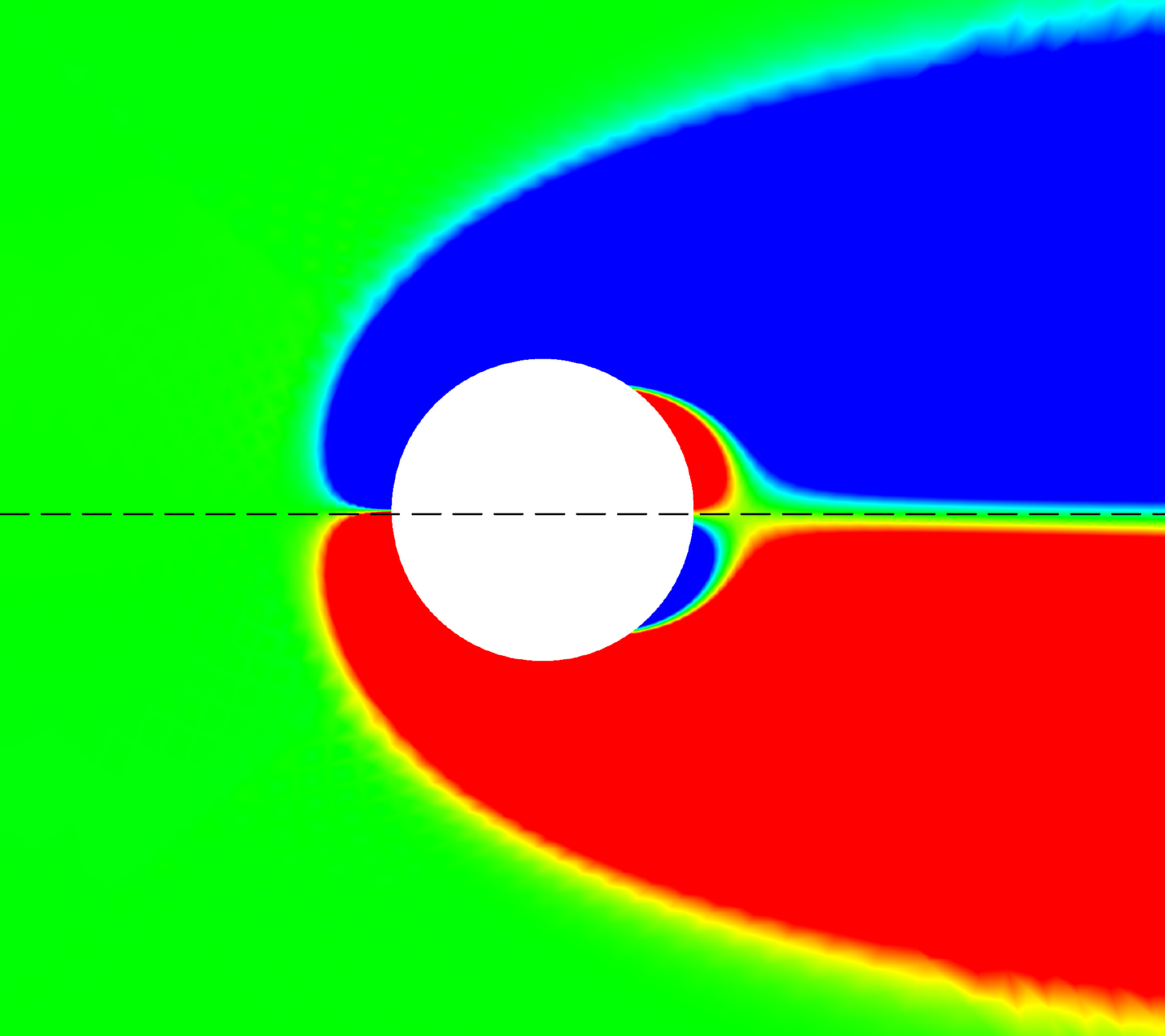}
\caption{$tU_{0}/D=65$}
\label{Re40-Ur5-t1point90}
\end{subfigure}
\begin{subfigure}{0.22\textwidth}
\centering
\includegraphics[width=0.9\linewidth]{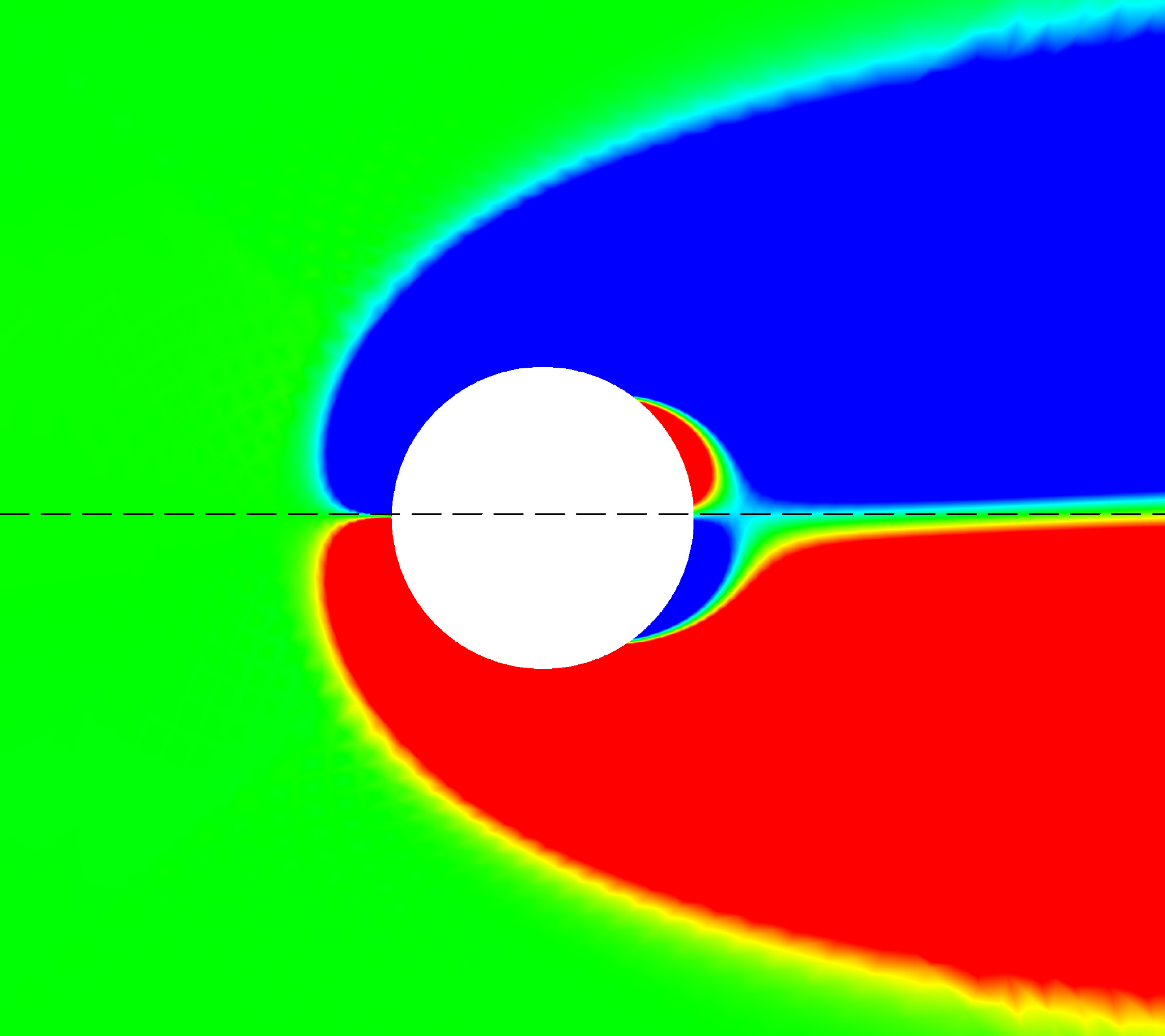}
\caption{$tU_{0}/D=70$}
\label{Re40-Ur5-t2point05}
\end{subfigure}%
\begin{subfigure}{0.22\textwidth}
\centering
\includegraphics[width=0.9\linewidth]{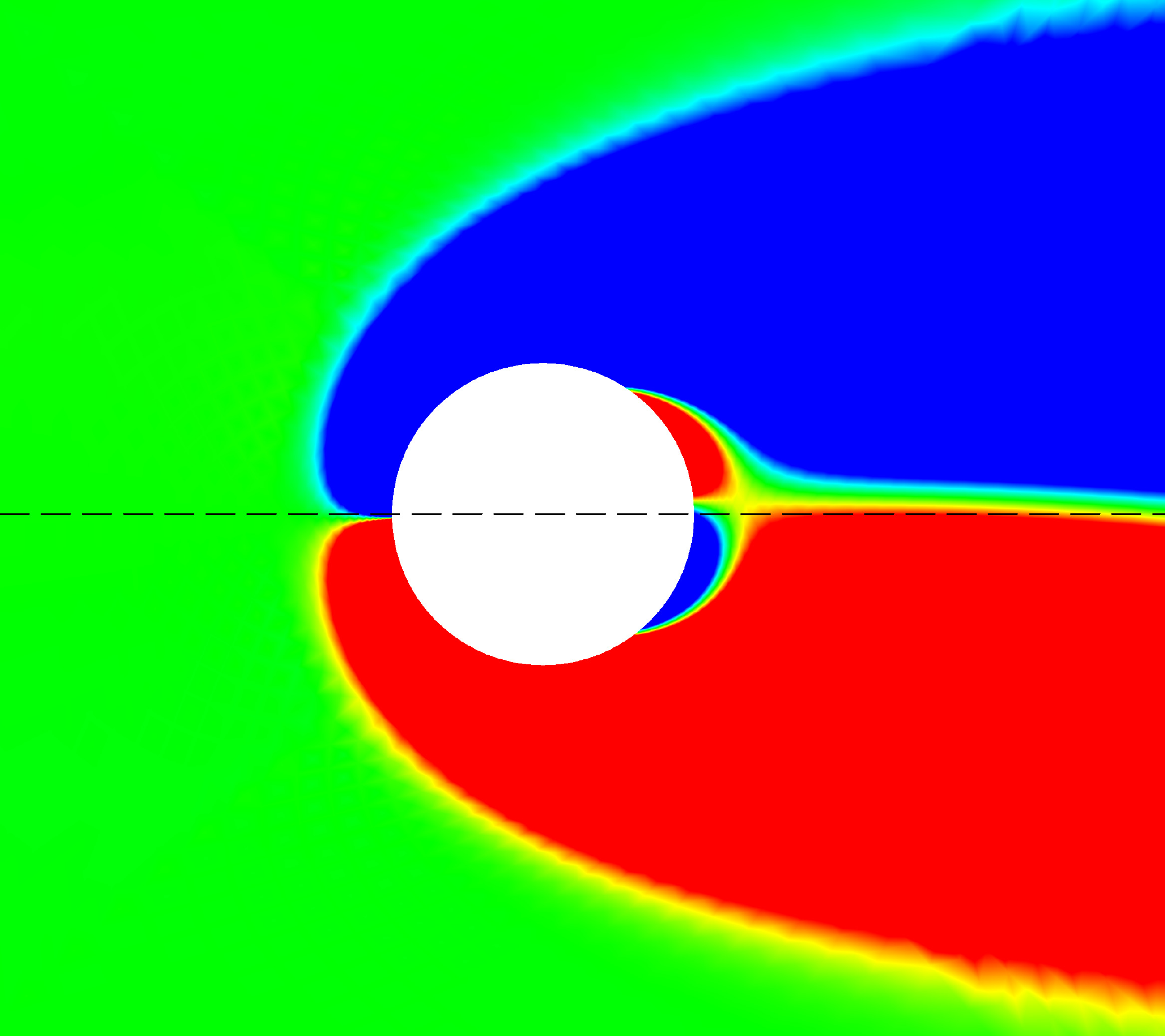}
\caption{$tU_{0}/D=75$}
\label{Re40-Ur5-t2point20}
\end{subfigure}
\begin{subfigure}{0.22\textwidth}
\centering
\includegraphics[width=0.9\linewidth]{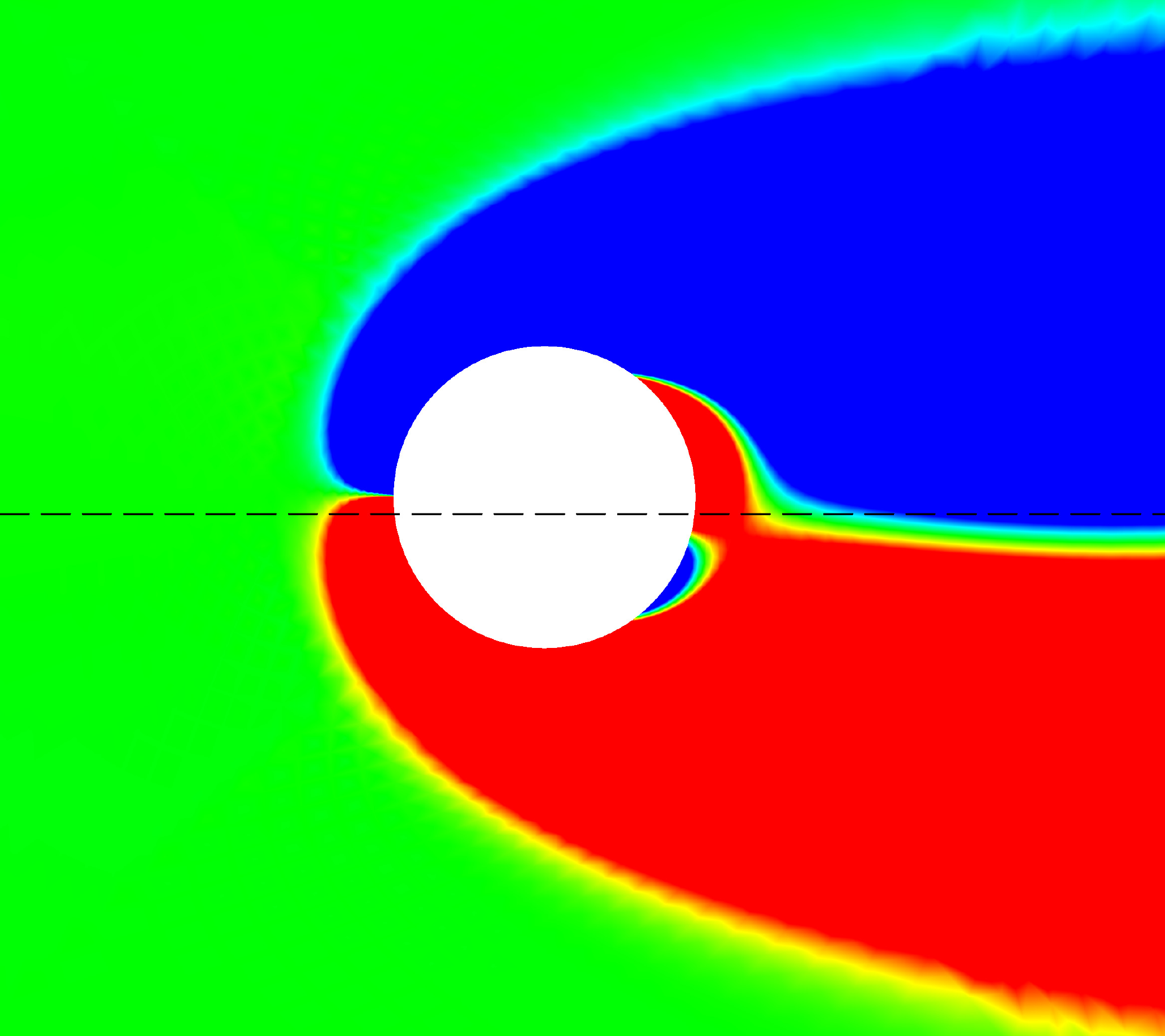}
\caption{$tU_{0}/D=80$}
\label{Re40-Ur5-t2point35}
\end{subfigure}%
\begin{subfigure}{0.22\textwidth}
\centering
\includegraphics[width=0.9\linewidth]{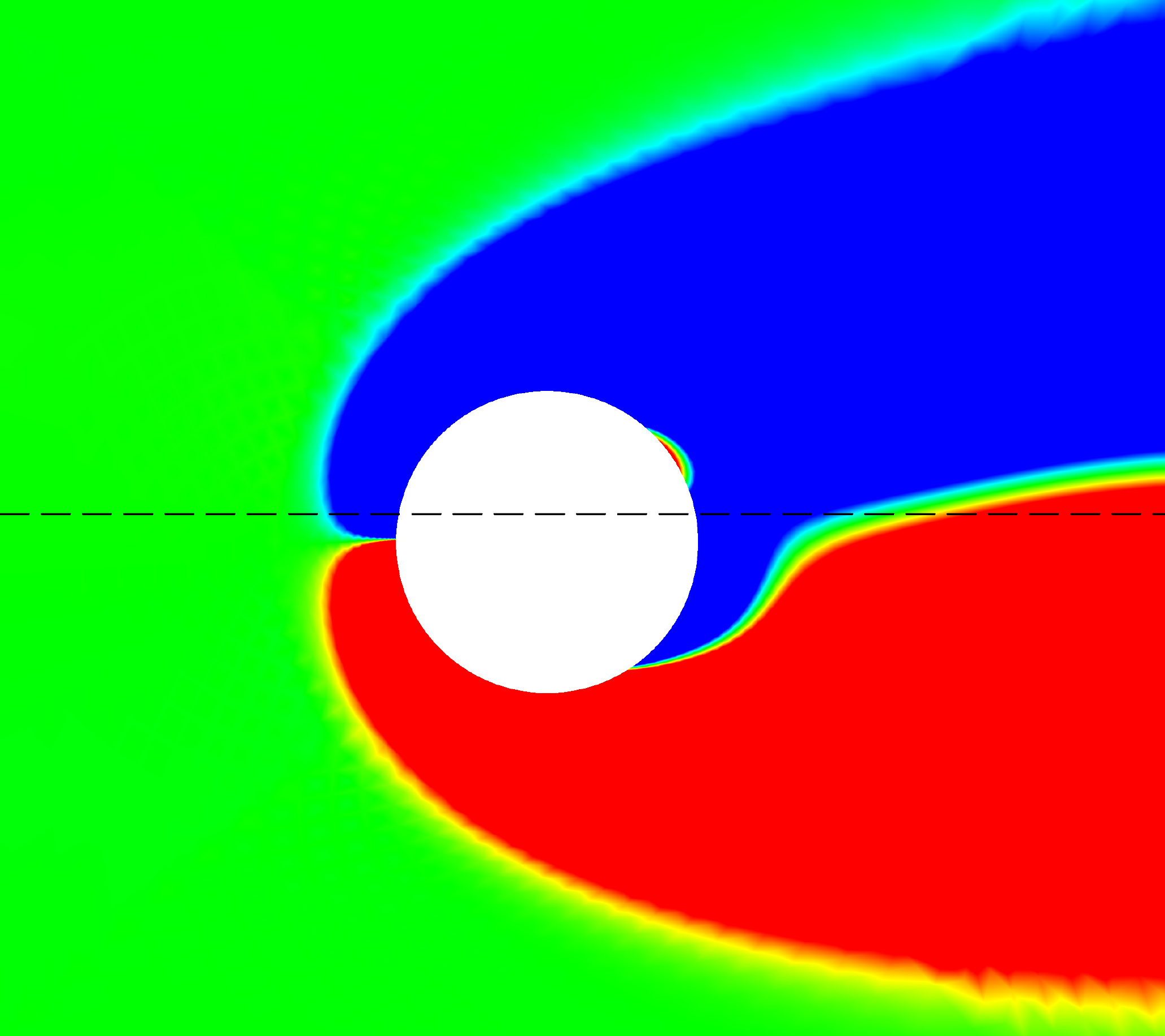}
\caption{$tU_{0}/D=85$}
\label{Re40-Ur5-t2point50}
\end{subfigure}
\begin{subfigure}{0.44\textwidth}
\centering
\includegraphics[width=0.95\linewidth]{Z-Vorticity-colorbar}
\end{subfigure}
\caption{\label{Re40-Ur5-zVor} Contours of z-vorticity at the cross section of the cylinder at $z/L=0.5$, $Re=40$, $m^*=1$, and $U^* = 7$ in the time window $tU_{0}/D\in[60, 85]$.}
\end{figure}

To better understand the relationship between the cylinder motion and stability of the wake at laminar subcritical $Re$, we have provided the z-vorticity contours at the mid-section of the cylinder at $Re=40$, $m^*=1$, and $U^*=7$ in Fig.~\ref{Re40-Ur5-zVor}. We show that the wake region behind the cylinder is steady and symmetric at $tU_{0}/D=60$; however, for $tU_{0}/D\in[65,75]$, relative motion of the cylinder cross-section in the transverse direction, makes the wake lose its stability and become asymmetric. This symmetry breakdown, in turn, exerts a transverse load that further increases the cylinder motion. Finally, due to the coupling between the unsteady wake and the cylinder movements, large-amplitude transverse vibrations are observed for $tU_{0}/D\in[80,85]$.

In the next section, we investigate the effect of mass ratio $m^*$ on the dynamics of the flexible cantilever cylinder and further examine the wake structures in the lock-in regime.
\subsection{Effect of mass ratio}
\begin{figure}[b]
\centering
\includegraphics[width=1\linewidth]{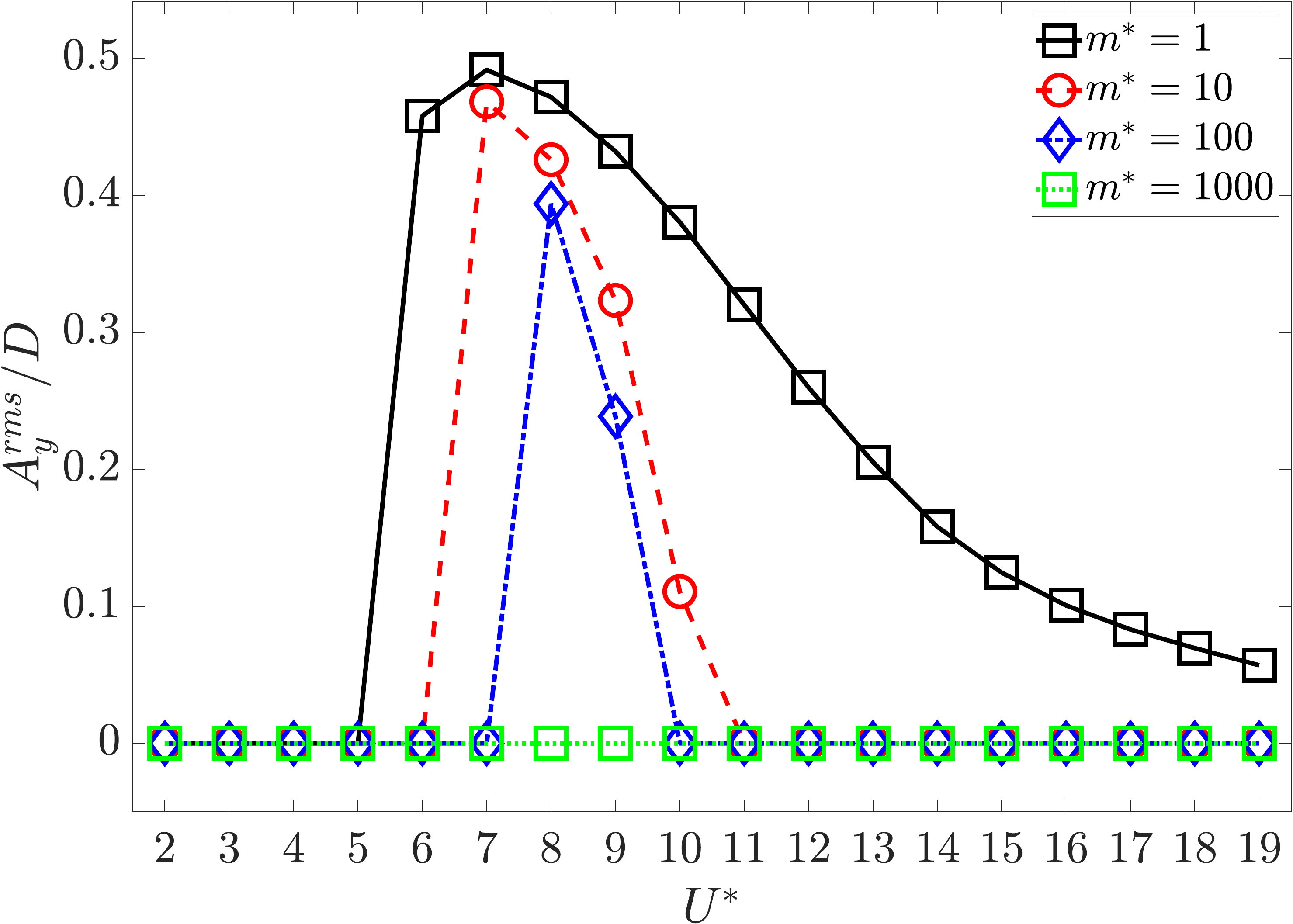}
\caption{\label{Ayrms-Ur-massratio}Root-mean-square value of the dimensionless transverse vibration amplitude $A_\mathrm{y}^{rms}/D$ at $z/L=1$ as a function of $U^*$ at $Re = 40$ for $m^*=1, 10, 100$ and $1000$.}
\end{figure}

We first investigate the effect of mass ratio on the dynamic response of the flexible cantilever cylinder at $Re=40$ for $U^*\in[2, 19]$. We examine the response of the cylinder at four different mass ratios, namely $m^* = 1, 10, 100$ and $1000$. The results for the rms value of the dimensionless transverse vibration amplitude $A_\mathrm{y}^{rms}/D$ with respect to $U^*$ are given in Fig.~\ref{Ayrms-Ur-massratio}. For all the studied mass ratios, we find that the cylinder stays at its steady deflected position for $U^*\leq5$. This steady response is present for the whole range of $U^*$ at $m^*=1000$. However, a discrete change in the dynamic response of the cylinder is observed for higher $U^*$ values at $m^* = 1, 10$ and $100$. We observe a sudden jump in the amplitude response of the cylinder at $U^*=6, 7$ and $8$ at mass ratios $m^*=1, 10$ and $100$, respectively. As shown in Fig.~\ref{Ayrms-Ur-massratio}, the peak of the transverse vibration amplitude is at $U^*=7$ for $m^*=1$ and $10$, and at $U^*=8$ for $m^*=100$. The magnitude of the maximum $A_\mathrm{y}^{rms}/D$ is shown to be approximately $0.49, 0.47$ and $0.39$ at $m^*=1, 10$ and $100$, respectively. By further increasing the $U^*$, a gradual decrease in the amplitude of the transverse vibrations is observed at $m^*=1$; however, for $m^*=10$ and $100$, there is a sharp decrease in the amplitude of the transverse vibrations for $U^*>8$. A steady response is observed for $U^*\geq11$ at $m^*=10$ and for $U^*\geq10$ at $m^*=100$. 

\begin{figure}
\includegraphics[width=1\linewidth]{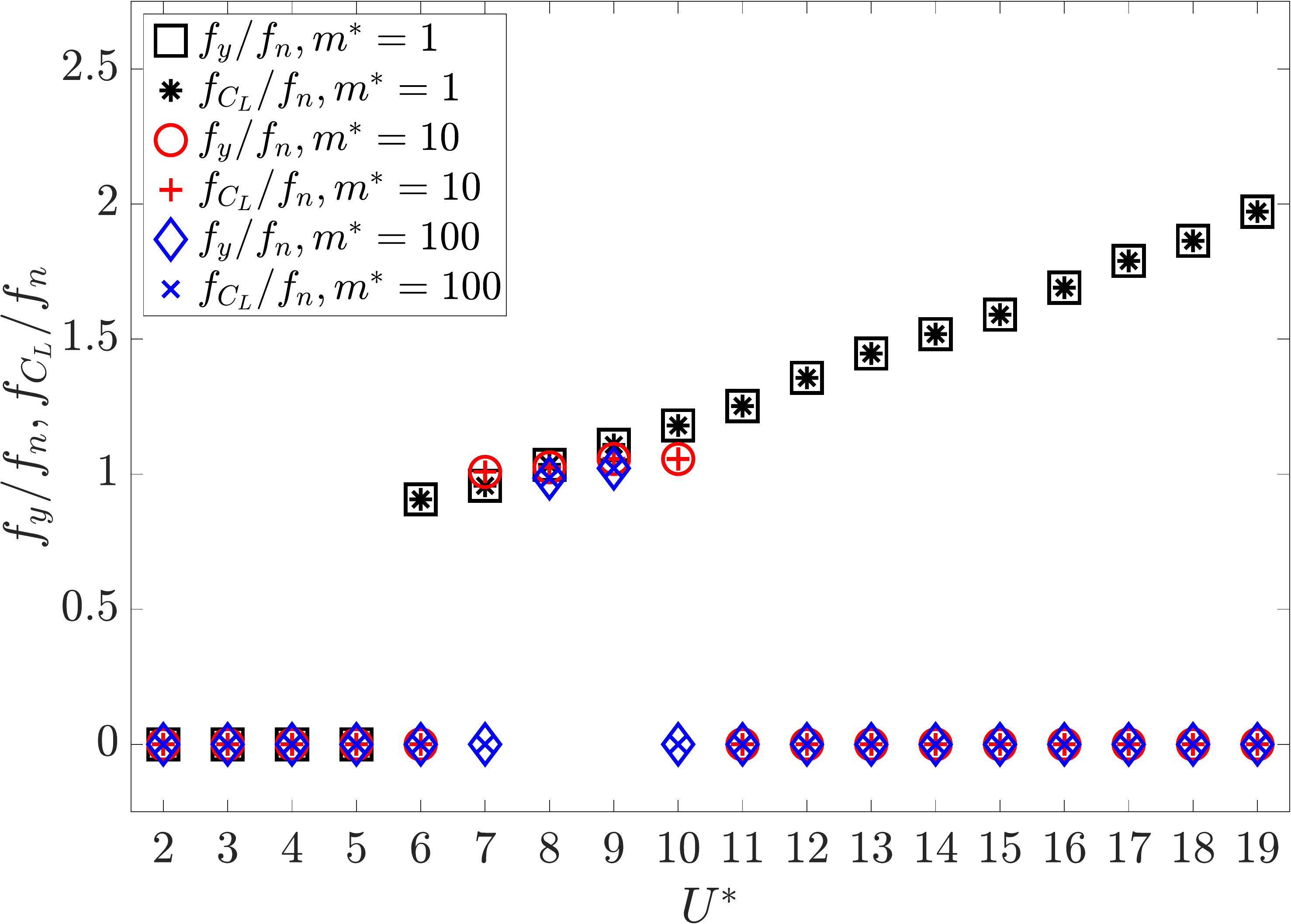}
\caption{\label{f_fn_Re40_mstar}Variations of the dimensionless transverse vibration frequency $f_\mathrm{y}/f_\mathrm{n}$, probed at $z/L=1$, and lift coefficient frequency $f_\mathrm{C_L}/f_\mathrm{n}$ with respect to $U^*$. The results are gathered at $Re=40$ for $m^* = 1, 10$ and $100$.}
\end{figure}

Fig.~\ref{f_fn_Re40_mstar} shows the frequency response of the fluid-structure system in terms of the dimensionless transverse vibration frequency $f_\mathrm{y}/f_\mathrm{n}$ and the dimensionless lift coefficient frequency $f_\mathrm{C_L}/f_\mathrm{n}$ at $Re=40$ for $m^*=1,10$ and $100$ with respect to $U^*$. We show that for all three mass ratios, there is a frequency match between the frequency of the transverse vibrations $f_\mathrm{y}$, frequency of the lift coefficient $f_\mathrm{C_L}$, and the first-mode natural frequency of the cylinder $f_\mathrm{n}$ for a specific range of $U^*$. This range is within $U^*\in[8,9]$ at $m^*=100$ and within $U^*\in[7,10]$ at $m^*=10$. At $m^*=1$, the lock-in regime begins at $U^*=6$, however, the beam is shown to oscillate in frequencies higher than its first-mode natural frequency at larger $U^*$ values.

Based on our findings, we observe that by increasing $m^*$, the range of the lock-in regime becomes narrower. This behavior is because of stronger inertial coupling and added mass effects at lower mass ratios. It should be noted that for Reynolds numbers beyond $Re_\mathrm{cr}\approx45$, interactions between the unsteady wake and the cylinder motion could lead to sustained vibrations for mass ratios of $O(100-1000)$, which are not examined in our current work. 

\begin{figure*}
\includegraphics[width=0.6\linewidth]{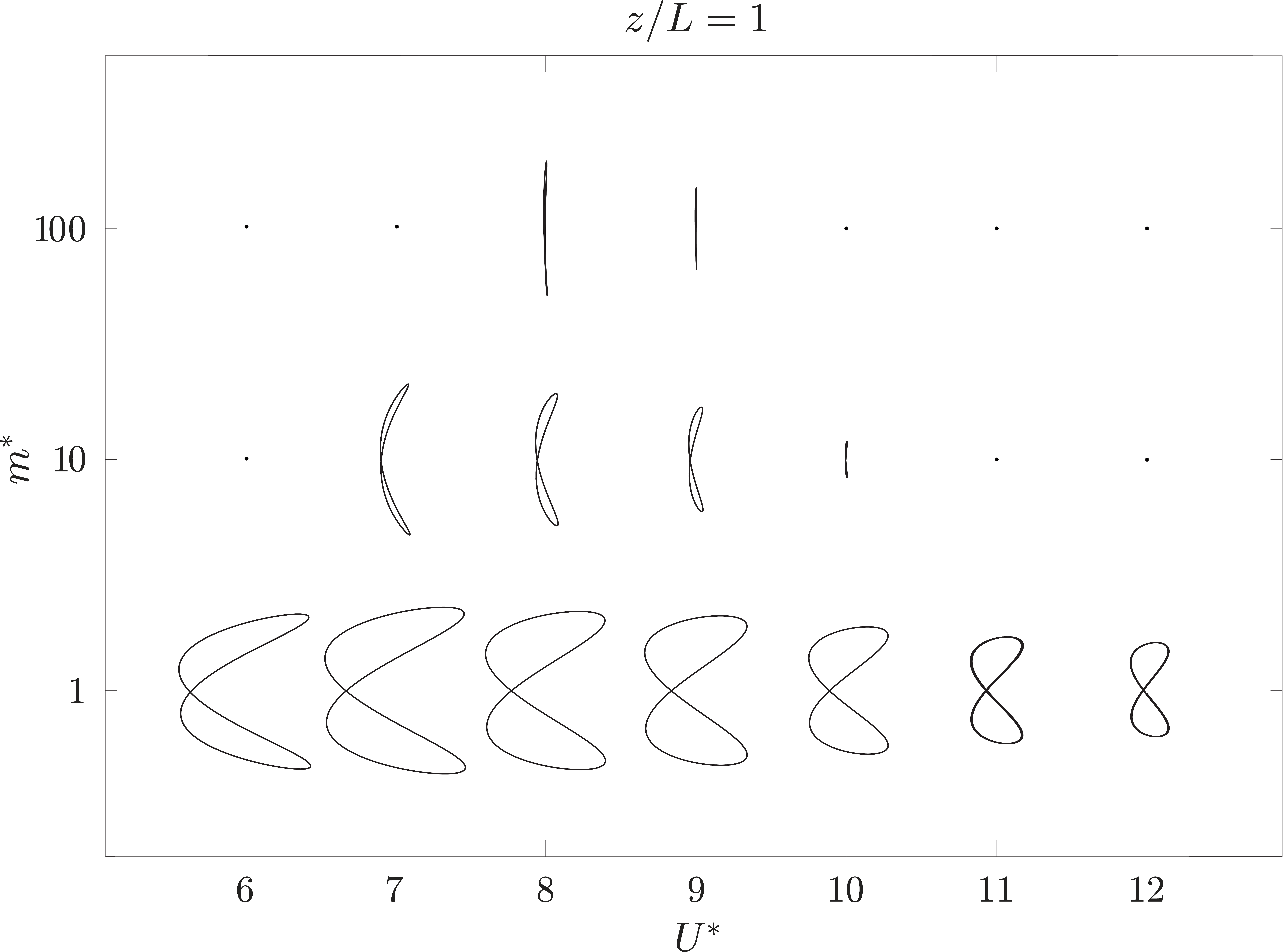}
\caption{\label{mstar-Ustar}A representation of the motion trajectory of the flexible cantilever cylinder with respect to $U^*$ at $z/L=1$ and $Re=40$ for $m^*=1, 10$ and $100$. The filled dot (.) represents a steady response.}
\end{figure*}

A qualitative representation of the cylinder motion trajectory at $z/L=1$ and $Re=40$ with respect to $U^*$ is given in Fig.~\ref{mstar-Ustar} for $m^*=1, 10,$ and $100$. We show that as $m^*$ is increased, the motion trajectory of the cylinder in the lock-in regime shifts from a figure-eight type response at $m^*=1$ to a dominated motion in the transverse direction at $m^*=100$.

\begin{figure*}
\begin{subfigure}{0.25\textwidth}
\centering
\includegraphics[width=0.7\linewidth]{z-vort-isosurface-Re40-Ur8-addedmass-mstar1}
\caption{$m^*=1$}
\label{isosurface-mstar1}
\end{subfigure}%
\begin{subfigure}{0.25\textwidth}
\centering
\includegraphics[width=0.7\linewidth]{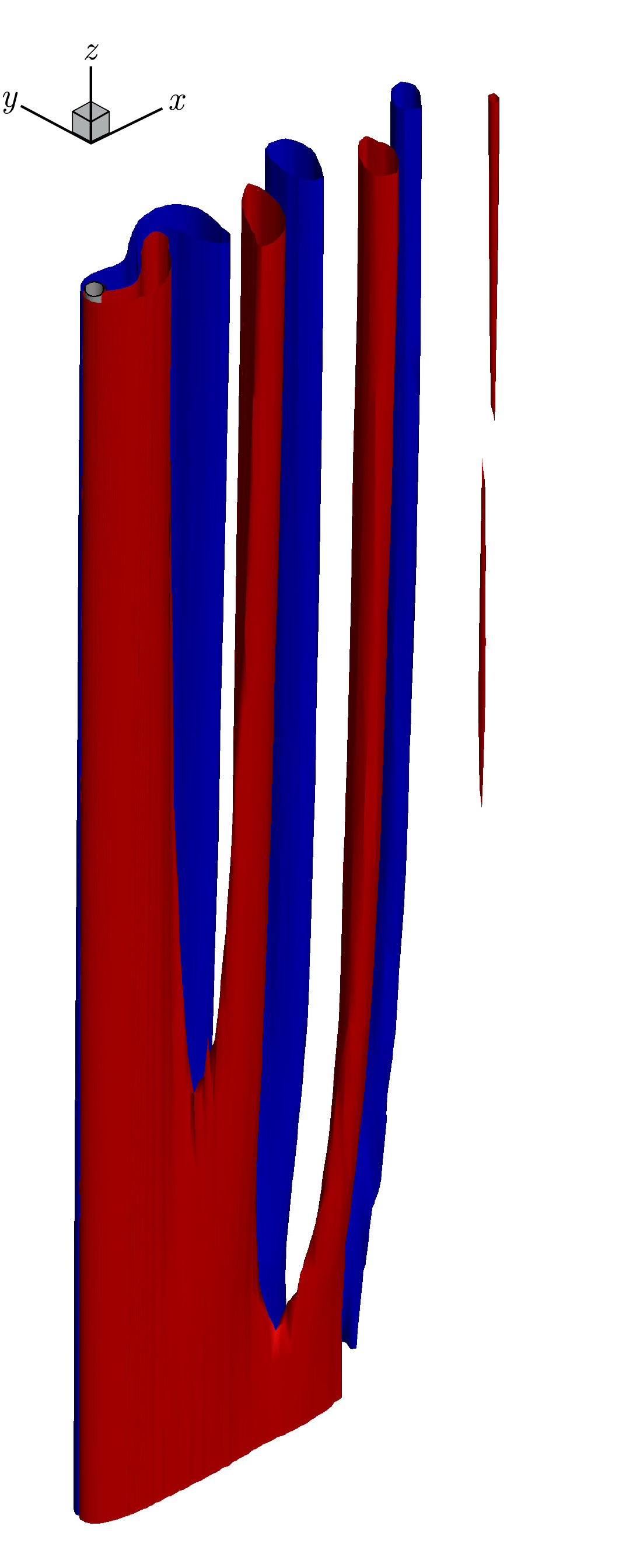}
\caption{$m^*=10$}
\label{isosurface-mstar10}
\end{subfigure}%
\begin{subfigure}{0.25\textwidth}
\centering
\includegraphics[width=0.7\linewidth]{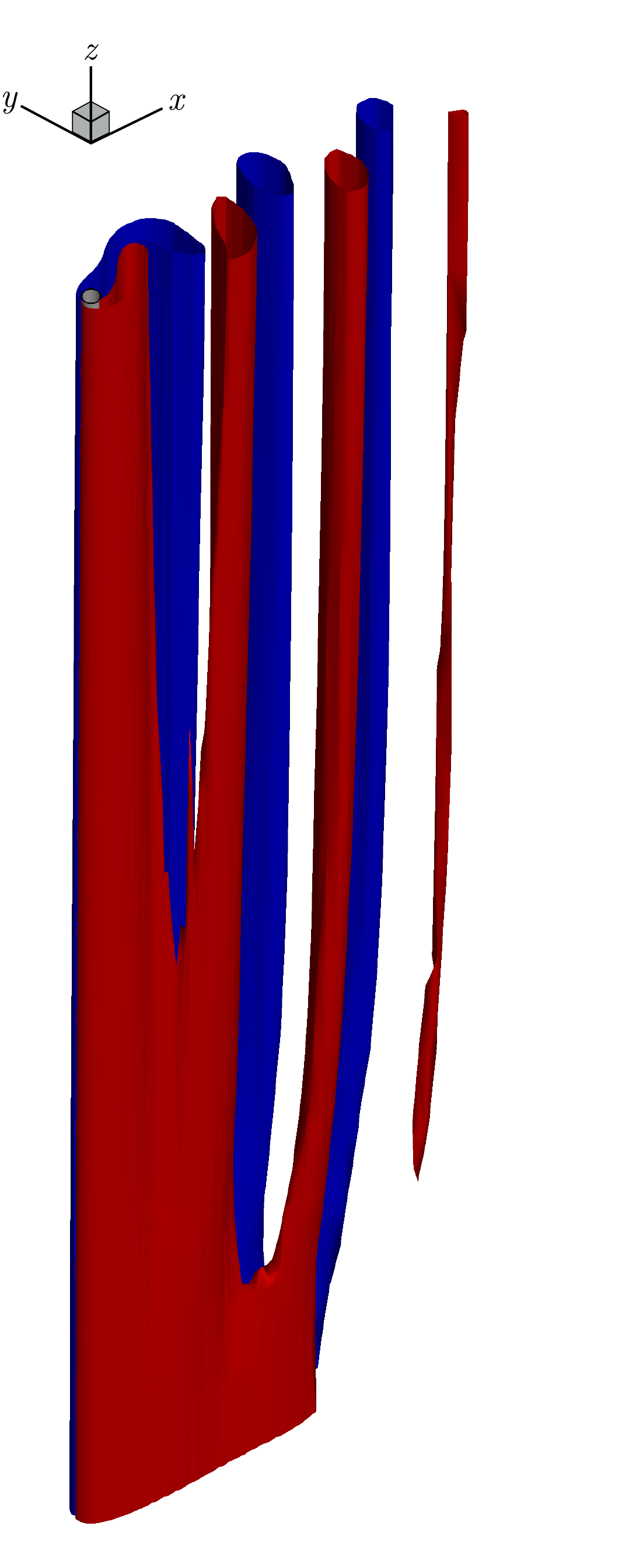}
\caption{$m^*=100$}
\label{isosurface-mstar100}
\end{subfigure}
\caption{\label{zvort-isosurfaces-ustar8-mstar}Wake structures visualized by the normalized z-vorticity iso-surfaces ($\omega_{z}D/U_{0} = -0.224,0.224$) for the flexible cantilever cylinder at $Re=40$, and $U^*=8$. Red [blue] indicates regions of positive [negative] vortices.}
\end{figure*}

To examine the wake structures in the lock-in regime at different mass ratios, we have provided the z-vorticity iso-surfaces around the cylinder at $Re=40$ and $U^*=8$ for $m^*=1,10$ and $100$ in Fig.~\ref{zvort-isosurfaces-ustar8-mstar}. We observe that for all three $m^*$, two alternate vortices are shed from the cylinder wake in each cycle. This finding suggests that at a fixed $Re$, the vortex shedding patterns in the wake of the flexible cantilever cylinder are independent of the mass ratio $m^*$.

Finally, we connect our findings to the dynamic response of rat and seal whiskers in fluid flow. The presented results in this section for $m^*=1$ and $100$ are qualitative representatives of a seal and a rat whisker in fluid flow, respectively. As mentioned in Section~\ref{sec:introduction}, the interaction of a rat's whisker with low-speed airflow occurs at $Re<50$. Based on our results for the flexible cantilever cylinder, a rat's whisker at laminar subcritical $Re$ is expected to experience a VIV-galloping type instability in the lock-in regime. In addition, flutter instability at reduced velocities of $O(100)$ could appear in a rat's whisker, which requires further investigations.

For the case of a seal whisker in water flow, we characterize the oscillations as a VIV-dominant mechanism that occurs due to interactions between the whisker and unsteady wake at $Re\approx1000$. Our results for the dynamic response of the flexible cantilever cylinder at $m^*=1$ could be used to interpret the response of a seal whisker in low-speed laminar water flows. 

Based on the available data in the literature regarding the amplitude response of an elastically mounted rigid cylinder at laminar $Re$~\cite{Bearman2011}, we can deduce that the amplitude response of the flexible cantilever cylinder at $Re\approx1000$ would be slightly higher than the values presented in our current work. In addition, the vibration amplitude of a seal whisker is predicted to be significantly lower than the vibration amplitude of the flexible cantilever cylinder under similar conditions. Based on the experiments by Refs~\cite{Beem2015,Hanke2010}, lower amplitudes in a seal whisker, compared to the flexible cantilever cylinder, are due to the seal whisker's undulated geometry that helps reduce the fluid forces during VIVs.

\section{\label{sec:conclusions} Conclusions}
In this paper, we have investigated the fluid-structure interaction of a flexible cantilever cylinder at laminar subcritical $Re$. Through numerical simulations, we assessed the dynamic response of the cylinder as a function of reduced velocity $U^*$, for Reynolds numbers between $20\leq Re\leq 40$ and mass ratios between $1\leq m^*\leq 1000$.
We found that for $Re=20$, the flexible cantilever cylinder remains in its steady deflected position for the whole range of studied $U^*$ and $m^*$. However, for $22\leq Re\leq 40$, we observed that the cylinder experiences sustained oscillations when certain conditions are satisfied. We showed that the frequency of the transverse vibrations matches the frequency of the periodic lift force during the oscillations. Also, these two frequencies were found to be approximately equal to the first-mode natural frequency of the cylinder for a particular range of $U^*$. This specific range, known as the lock-in regime, was shown to strongly depend on the Reynolds number $Re$ and mass ratio $m^*$; we found that at laminar subcritical $Re$, the range of the lock-in regime decreases by increasing $m^*$, whereas this range was shown to increase by increasing $Re$. Finally, we identified two requirements for the wake unsteadiness at laminar subcritical $Re$: (i) the flow needs to have sufficiently large inertia to overcome the viscous damping, and (ii) the system parameters need to be in the lock-in range. When these two conditions are satisfied, the cylinder experiences a combined VIV-galloping type instability. This instability was shown to break the symmetry of the wake and lead to sustained large-amplitude vibrations and unsteadiness in the wake at laminar subcritical $Re$.
We anticipate that the presented systematic analysis can help improve our understanding of the lock-in mechanism in flexible cantilever structures. Further research is required towards a parametric investigation of the dynamic response of the cylinder at high reduced velocities of $O(100)$, where potential flutter-type instabilities could be present. In addition, the effects of structural nonlinearities should be accounted from a practical viewpoint for different flow incidence with a broader range of $Re$ and $U^*$ to fully understand the dynamic instabilities of the coupled system. 
\begin{acknowledgments}
The authors would like to acknowledge the Natural Sciences and Engineering Research Council of Canada (NSERC) for the funding. This research was enabled in part through computational resources and services provided by WestGrid (\url{https://westgrid.ca/}), Compute Canada (\url{https://computecanada.ca/}), and the Advanced Research Computing facility at the University of British Columbia (\url{https://arc.ubc.ca/}).
\end{acknowledgments}


\bibliography{main}

\end{document}